\documentclass[a4paper,twocolumn,%tightenlines,
english,aps,pra,floatfix,showpacs]{revtex4}
\usepackage[latin1]{inputenc}
\usepackage{amsmath}
\usepackage{babel}
\usepackage{graphics}
\usepackage{amssymb}



\def\s{\sigma}

\begin{document}

\newcommand{\dt}{\Delta\tau}
\newcommand{\al}{\alpha}
\newcommand{\ep}{\varepsilon}
\newcommand{\ave}[1]{\langle #1\rangle}
\newcommand{\have}[1]{\langle #1\rangle_{\{s\}}}
\newcommand{\bave}[1]{\big\langle #1\big\rangle}
\newcommand{\Bave}[1]{\Big\langle #1\Big\rangle}
\newcommand{\dave}[1]{\langle\langle #1\rangle\rangle}
\newcommand{\bigdave}[1]{\big\langle\big\langle #1\big\rangle\big\rangle}
\newcommand{\Bigdave}[1]{\Big\langle\Big\langle #1\Big\rangle\Big\rangle}
\newcommand{\braket}[2]{\langle #1|#2\rangle}
\newcommand{\up}{\uparrow}
\newcommand{\dn}{\downarrow}
\newcommand{\bb}{\mathsf{B}}
\newcommand{\ctr}{{\text{\Large${\mathcal T}r$}}}
\newcommand{\sctr}{{\mathcal{T}}\!r \,}
\newcommand{\btr}{\underset{\{s\}}{\text{\Large\rm Tr}}}
\newcommand{\lvec}[1]{\mathbf{#1}}
\newcommand{\gt}{\tilde{g}}
\newcommand{\ggt}{\tilde{G}}
\newcommand{\jpsj}{J.\ Phys.\ Soc.\ Japan\ }

\centerline{Braz.\ J.\ Phys.\ \bf{33}, 36 (2003)}
 
\title{Introduction to Quantum Monte Carlo simulations for 
fermionic systems} 

\author{Raimundo R.\ \surname{dos Santos}}

\email{rrds@if.ufrj.br}

\affiliation{Instituto de F\'\i sica, Universidade Federal do
Rio de Janeiro, Caixa Postal 68528, 21945-970
Rio de Janeiro RJ, Brazil}

\date{\today}

\begin{abstract}
We tutorially review the determinantal Quantum Monte 
Carlo method for fermionic systems, using the Hubbard model as a case
study. Starting with the basic ingredients of Monte Carlo simulations 
for classical systems, we introduce aspects such as importance sampling, 
sources of errors, and finite-size scaling analyses. We then set up the 
preliminary steps to prepare for the simulations, showing that they are
actually carried out by sampling discrete Hubbard-Stratonovich auxiliary 
fields. In this process the Green's function emerges as a fundamental 
tool, since it is used in the updating process, and, at the same time, 
it is directly related to the quantities probing magnetic, charge, 
metallic, and superconducting behaviours. We also discuss the 
as yet unresolved `minus-sign problem', and two ways to stabilize the
algorithm at low temperatures.
\end{abstract}
\maketitle
%\tightenlines
\section{Introduction}
\label{intro}

In dealing with systems of many interacting fermions, one is generally
interested in their collective properties, which are suitably described
within the framework of Statistical Mechanics. Unlike insulating magnets,
in which the spin degrees of freedom are singled out, the interplay
between charge and spin is responsible for a wealth of interesting
phenomena (orbital degrees of freedom may also be included, but they add
enormously to complexity, and shall not be considered here). Typical
questions one asks about a system are related to its magnetic state (Is it
magnetic? If so, what is the arrangement?), to its charge distribution,
and to whether it is insulating, metallic, or superconductor.

A deeper understanding of the interplay between spin and charge degrees of
freedom can be achieved through {\it models}, which, while capturing the
basic physical mechanisms responsible for the observed behaviour, should
be simple enough to allow calculations of quantities comparable with
experiments. The simplest model describing interacting fermions on a
lattice is the single band Hubbard model\cite{Hub63}, defined by the
grand-canonical Hamiltonian
\begin{eqnarray}
&{\cal H}=&-t\sum_{\langle{\bf i},{\bf j}\rangle,\,\sigma} 
 \left(c_{{\bf i}\sigma}^{\dagger}c_{{\bf j}\sigma}^{\phantom{\dagger}}+ 
{\rm 
H.c.}\right)+
         U\sum_{\bf i} n_{{\bf i}\uparrow}^{\phantom{\dagger}} 
         n_{{\bf i}\downarrow}^{\phantom{\dagger}}-\nonumber\\
&&-\mu \sum_{\bf i} 
\left(n_{{\bf i}\uparrow}^{\phantom{\dagger}}+
      n_{{\bf i}\downarrow}^{\phantom{\dagger}}\right), \label{Hub}
\end{eqnarray} 
where $t$ is the hopping integral (which sets the energy scale, so we take
$t=1$ throughout this paper), $U$ is the on-site Coulomb repulsion, $\mu$
is the chemical potential controlling the fermion density, and {\bf i}
runs over the sites of a $d$-dimensional lattice; for the time being, 
we consider hopping
between nearest neighbours only, as denoted by $\langle\ldots\rangle$. The
operators $c_{{\bf i}\sigma}^{\dagger}$ and $c_{{\bf
i}\sigma}^{\phantom{\dagger}}$ respectively create and annihilate a
fermion with spin $\sigma$ on the (single) orbital centred at ${\bf i}$,
while $n_{{\bf i}\sigma}\equiv c_{{\bf i}\sigma}^{\dagger}c_{{\bf
i}\sigma}^{\phantom{\dagger}}$. The Hubbard model describes the
competition between opposing tendencies of itinerancy (driven by the
hopping term), and localization (driven by the on-site repulsion). For a
half-filled band (one fermion per site), it can be shown\cite{Emery76}
that in the limit of strong repulsion the Hubbard Hamiltonian reduces to
that of an isotropic antiferromagnetic Heisenberg model with an exchange
coupling $J=4t^2/U$.

If we let the coupling $U$ assume negative values, one has the
\emph{attractive} Hubbard model. Physically, this \emph{local} attraction
can have its origin in the coupling of fermions to extended (through
polaron formation)  or local phonons (such as vibrational modes of
chemical complexes)  \cite{MMR90}. Given that its weak coupling
limit describes the usual BCS theory, and that the real-space pairing
is more amenable to numerical calculations, this model has also been
useful in elucidating several properties of both conventional and
high-temperature superconductors \cite{MMR90}.

The Hubbard model [even in its simplest form, Eq.~(\ref{Hub})] is only
exactly soluble in one dimension, through the Bethe ansatz; correlation
functions, however, are not directly available. In higher dimensions one
has to resort to approximation schemes, and numerical techniques such as
Quantum Monte Carlo (QMC) simulations have proven to be crucial in
extracting information about strongly correlated fermions. 

Since the first Monte Carlo method for classical systems was devised in
the early 1950's \cite{Metro,Koonin,Binder}, several QMC algorithms have
been proposed. Their classification is varied, and depends on which aspect
one wishes to single out. For instance, they can be classified according
to whether the degrees of freedom lie in the continuum or on a lattice; or
whether it is a ground-state or a finite-temperature framework; or whether
it is variational or projective; or even according to some detailed aspect of
their implementation, such as if an auxiliary field is introduced, or if a
Green's function is constructed by the power method. Excellent broad
overviews of these algorithms are available in the literature, such as
Refs.\ \cite{Suzuki,vdl92}, so here we concentrate on the actual details
of the grand-canonical formulation with auxiliary fields leading to
fermionic determinants \cite{BSS}. We will pay special attention to the
implementations and improvements achieved over the years
\cite{Hirsch83,Hirsch85,Hirsch88,White89,Loh92}. We will also have in mind
primarily the Hubbard model, but will not discuss at length the results
obtained; instead, illustrative references will be given to guide the
reader to more detailed analyses, and we apologize for the omitions of
many relevant papers, which was dictated by the need to keep the discussion 
focused on this particular QMC implementation, and not on 
the Hubbard (or any other) model.

In line with the tutorial purpose of this review, we introduce in Sec.\
\ref{cl-mc} the basic ingredients of Monte Carlo simulations, illustrated
for `classical' spins. In this way, we have the opportunity to draw the
attention of the inexperienced reader to the importance of thorough data
analyses, common to both classical and quantum systems, before embarking
into the more ellaborate quantum formalism. Preliminary manipulations,
approximations involved, and the natural appearance of Green's functions
in the framework are then discussed in Sec.\ \ref{qmc-p}. In Sec.\
\ref{qmc-s}, we describe the updating process, as well as the wide range
of average quantities available to probe different physical properties of
the systems. We then address two of the main difficulties present in the
simpler algorithm presented so far: the still unresolved `minus-sign
problem' (Sec.\ \ref{minus}), and the instabilities at low-temperatures,
(Sec.\ \ref{stable}) for which two successful solutions are discussed. 
Conclusions and some perspectives are then presented in Sec.\
\ref{conc}.

\section{Monte Carlo simulations for `classical' spins}
\label{cl-mc}

Our aim is to calculate quantities such as the partition function (or the 
grand-partition function) and various averages, including
correlation functions; these quantities are obtained by summing over all
configurations of the system.  For definiteness, let us consider the case
of the Ising model, 
\begin{equation}
{\cal H}= -J\sum_{\langle {\bf i},{\bf j}\rangle} 
\s^z_{\bf i}\s^z_{\bf j},
\label{Ising}
\end{equation}
where $J$ is the exchange coupling and $\sigma_{\bf i}^z=\pm 1$.

For a lattice with $N_s$ sites there are $2^{N_s}$ possible configurations
in phase space. Clearly all these configurations are not equally
important: recall that the probability of occurrence of a given
configuration, $C$, with energy $E(C)$, is $\propto
\exp{\left[-E(C)/k_BT\right]}$. Therefore, from the numerical point of
view, it is not efficient to waste time generating {\it all}
configurations. Accordingly, the basic Monte Carlo strategy consists of an
{\it importance sampling} of configurations\cite{Koonin,Binder}. This, in turn,
can be implemented with the aid of the {\it Metropolis algorithm\
}\cite{Metro}, which generates, in succession, a chain of the most likely
configurations (plus fluctuations; see below). Starting from a random spin
configuration $C \equiv |\s_{\bf 1}, \s_{\bf 2},\ldots,\s_{\bf
N_s}\rangle$, one can imagine a walker visiting every lattice site and
attempting to flip the spin on that site.  To see how this is done, let us
suppose the walker is currently on site ${\bf i}$, and call $C'$ the
configuration obtained from $C$ by flipping $\s_{\bf i}$. These
configurations differ in energy by $\Delta E=E(C')-E(C)=2J\s_{\bf
i}\sum_{\bf j}^\prime \s_{\bf j}$, where the prime in the sum
restricts ${\bf j}$ to nearest neighbour sites of ${\bf i}$. The ratio
between the corresponding Boltzmann factors is then
\begin{equation}
r'\equiv \frac{p(C')}{p(C)} = \exp{\left(-\Delta E/k_BT\right)}.
\label{Boltz}
\end{equation}
Thus, if $\Delta E < 0$, $C'$ should be accepted as a new configuration in
the chain. On the other hand, if $\Delta E > 0$ the new configuration is
less likely, but can still be accepted with probability $r'$; this
possibility simulates the effect of fluctuations. 
Alternatively\cite{Binder}, one
may adopt the \emph{heat-bath algorithm,} in which the configuration
$C'$ is accepted with probability 
\begin{equation}
r\equiv \frac{r'}{1+r'}
\label{heatbath}
\end{equation}
In both cases, the local character of the updating of the configuration 
should be noted;  that is, the acceptance of the flip does not influence 
the state of all other spins. As we will see, this is not true for quantum 
systems. 

The walker then moves on, and attempts to flip the spin on the next site 
through one of the processes just described. 
After sweeping through the whole lattice, the 
walker goes back to the first site, and starts attempting to flip spins 
on all sites again.
One should make sure that the walker sweeps through the lattice many times 
so that the system thermalizes at the given temperature; this \emph{warming-up} stage  takes typically between a few hundred and a few 
thousand sweeps, but it can be very slow in some cases. 

Once the system has warmed-up, we can start `measuring' average values.  
Suppose that at the end of the $\zeta$-th sweep we have stored a value
$A_\zeta$ for the quantity $A$. It would therefore seem natural that after
$N_a$ such sweeps, an estimate for the thermodynamic average $\ave{A}$
should be given by
\begin{equation}
{\bar A}= \frac{1}{N_a} \sum_{\zeta=1}^{N_a} A_\zeta.
\label{avebar}
\end{equation}
However, the $A_\zeta$'s are not independent random variables, since, by 
construction, the configurations in the chain are somewhat correlated.
One can decrease correlations between measurements by forming 
a group of $G$ averages, ${\bar A}_1,{\bar A}_2,\ldots,{\bar A}_G$,
leading to a final average
\begin{equation}
\ave{A}= \frac{1}{G} \sum_{g=1}^G {\bar A}_g.
\label{ave}
\end{equation}
Ideally, one should alternate $N_n$ sweeps without taking 
measurements 
with $N_a$ sweeps in which measurements are taken. 

Once the $A_g$'s can be considered as independent random variables,
the central limit theorem \cite{Reif} applies, and the {\it statistical
errors} in the average of $A$ are estimated as
\begin{equation}
\delta A= \left[\frac{\ave{A^2}-\ave{A}^2}{G}\right]^{1/2}. 
\label{error}
\end{equation}

In addition, {\it systematic errors} must be taken into consideration, the
most notable of which are finite-size effects. Indeed, one should 
have in mind that usually there are two important length scales in these 
calculations: the linear size, $L$, and the correlation length
of the otherwise infinite system, $\xi\sim |T-T_c|^{-\nu}$. A 
generic thermodynamic quantity, $X_L(T)$, then scales as\cite{fss}
\begin{equation}
X_L(T)=L^{x/\nu}f\left(L/\xi\right),
\label{fss1}
\end{equation}
where $x$ is a critical exponent to be defined below, and $f(z)$ is a
scaling function with very specific behaviours in the limits $z\ll 1$ and
$z\gg 1$. In the former limit it should reflect the fact that the range of
correlations is limited by the finite system size, and one must have
\begin{equation}
X_L(T)\simeq {\rm const.}\cdot L^{x/\nu},\ {\rm for}\ L\ll \xi,
\label{fss2}
\end{equation}
so that an explicit $L$-dependence appears. On the other hand, when
correlations do not detect the finiteness of the system, the scaling
function must restore the usual size-independent form,
\begin{equation}
X_L(T)\simeq |T-T_c|^{-x},\ {\rm for}\ L\gg \xi;
\label{fss3}
\end{equation}
this defines the critical exponent $x$. This finite-size scaling
(FSS) theory can be used in the analysis of data, thus setting
the extrapolation towards the thermodynamic limit on firmer grounds.

As a final remark, we should mention for completeness that the actual
implementation of the Monte Carlo method for classical spins can be 
optimized in several aspects, such as using bit-strings to represent
states \cite{pmco-book}, and broad histograms to collect 
data \cite{pmco-broad}.

\section{Determinantal Quantum Monte Carlo: Preliminaries}
\label{qmc-p}

The above discussion on the Ising model was tremendously simplified due to
the fact that the eigenstates of the Hamiltonian are given as products
over single particle states. Quantum effects manifest themselves in the
fact that different terms in the Hamiltonian do not commute. In the case
of the Hubbard model, the interaction and hopping terms do not commute
with each other, and, in addition, hopping terms involving the same site
also do not commute with each other.  Consequently, the particles lose
their individuality since they are correlated in a fundamental way.

One way to overcome these non-commuting aspects \cite{BSS} is to notice
that the Hamiltonian contains terms which are bilinear in fermion
operators (namely, the hopping and chemical potential terms) and a term
with four operators (the interaction term). Terms in the former category
can be trivially diagonalized, but not those in the latter. Therefore,
when calculating the grand partition function,
\begin{equation}
{\cal Z}= {\cal T}r\ e^{-\beta {\cal H}},
\label{z1}
\end{equation}
where, as usual, ${\cal T}r$ stands for a sum over all numbers of 
particles and over all site occupations, one must 
cast the quartic term into a bilinear form.

To this end, we first separate the exponentials with the aid of the
Suzuki-Trotter decomposition scheme\cite{Suzuki2}, which is based on the
fact that
\begin{equation}
e^{\Delta\tau (A+B)}= e^{\Delta\tau A}e^{\Delta\tau B} + 
{\cal O}\left[(\Delta\tau)^2\left[A,B\right]\right],
\label{Trotter}
\end{equation}
for $A$ and $B$ generic non-commuting operators. Calling ${\cal K}$ and
${\cal V}$, respectively the bilinear and quartic terms in the Hubbard
Hamiltonian, we introduce a small parameter $\Delta\tau$ through
$\beta=M\, \Delta\tau$, and apply the Suzuki-Trotter formula as
%\begin{widetext}
\begin{eqnarray}
&e^{-\beta({\cal K}+{\cal V})}&=\left(e^{\Delta\tau{\cal 
K}+\Delta\tau{\cal V}}\right)^M=\nonumber\\
&&=\left(e^{\Delta\tau{\cal K}}e^{\Delta\tau{\cal V}}\right)^M
+{\cal O}\left[(\Delta\tau)^2 U\right].
\label{im-time}
\end{eqnarray}
%\end{widetext}
The analogy with the path integral formulation of Quantum Mechanics
suggests that the above procedure amounts to the \emph{imaginary-time}
interval $(0,\beta)$ being discretized into $M$ slices separated by the
interval $\Delta\tau$.  

In actual calculations at a fixed inverse temperature $\beta$, we
typically set $\Delta\tau= \sqrt{0.125/U}$ and choose 
$M=\beta/\Delta\tau$.  As
evident from Eq.\ (\ref{im-time}), the finiteness of $\Delta\tau$ is also
a source of {\it systematic errors;} these errors can be downsized by
obtaining estimates for successively smaller values of $\Delta\tau$ (thus
increasing the number, $M$, of time slices) and extrapolating the results
to $\Delta\tau\to 0$. Given that this complete procedure can be very
time consuming, one should at least check that the results are not too
sensitive to $\Delta\tau$ by performing calculations for two values of
$\Delta\tau$ and comparing the outcomes.

Having separated the exponentials, we can now workout the quartic terms 
in $\mathcal V$. 
We first recall a well known trick to change the exponential of the square 
of an operator into an exponential of the operator itself, 
known as the Hubbard-Stratonovich (HS) transformation,
\begin{equation}
e^{\frac{1}{2}A^2}\equiv\sqrt{2\pi}\int_{-\infty}^{\infty} 
dx\ e^{-\frac{1}{2}x^2-xA},
\label{HSc}
\end{equation}
at the price of introducing an auxiliary degree of freedom (field) $x$,
which couples linearly to the original operator $A$;  this result is
immediately obtained by completing the squares in the integrand. However,
before a transformation of this kind can be applied to the quartic term of
the Hubbard Hamiltonian, squares of operators must appear in the argument
of the exponentials. Since for fermions one has 
$n_\s^2\equiv n_\s=0,1$ (here we omit lattice indices to simplify the 
notation), the following identities -- in terms of the local 
magnetization, $m\equiv n_\uparrow - n_\downarrow$, and of the local 
charge, $n\equiv n_\uparrow + n_\downarrow$ -- will suit our purposes:
\begin{subequations}
\label{id}
\begin{eqnarray}
n_\uparrow n_\downarrow&=&-\frac{1}{2}m^2 + \frac{1}{2}n,
\label{id1}
\\
n_\uparrow n_\downarrow&=&\frac{1}{2}n^2 - \frac{1}{2}n,
\label{id2}
\\
n_\uparrow n_\downarrow&=&\frac{1}{4}n^2 - \frac{1}{4}m^2,\label{id3}
\end{eqnarray}
\end{subequations}
The following points are worth making about Eqs.\ (\ref{id}), in relation
to the HS transformation: (i) an auxiliary field will be introduced for
each squared operator appearing on the RHS's above; (ii) the auxiliary
field will couple to the local magnetization and to the local charge when,
respectively, Eqs.\ (\ref{id1}) and (\ref{id2}) are  used; (iii) 
Eqs.\ (\ref{id1}) and (\ref{id2}) are respectively used for the repulsive 
and for the attractive models.

Instead of the continuous auxiliary field of Eq.\ (\ref{HSc}), in
simulations it is more convenient to work with \emph{discrete} 
Ising variables, $s=\pm 1$ \cite{Hirsch83}. Inspired by Eq.\ (\ref{HSc}),
using Eq.\ (\ref{id1}), and taking into account that $s^2\equiv 1$, it is
straightforward to see that
\begin{eqnarray}
&e^{-U\dt n_\up n_\dn}&=\frac{1}{2} e^{-\frac{U\dt}{2} n} \sum_{s=\pm 1} 
e^{-s\lambda m}=\nonumber \\
&&=\frac{1}{2}\sum_{s=\pm 1}\prod_{\s=\up,\dn}e^{-\left(\s s \lambda + 
\frac{U\dt}{2}\right)n_\s},\ \ U>0,\nonumber\\
\label{HSdr}
\end{eqnarray}
where the grouping in the last equality factorizes the 
contributions from the two fermion spin 
channels, $\s=+,-$ (respectively corresponding to $\s=\up,\dn$), and
\begin{equation}
\cosh\lambda = e^{|U|\dt/2}.
\label{lamb}
\end{equation}

In the attractive case, the coupling to the charge [Eq.\ (\ref{id2})] is
used in order to avoid a complex HS transformation; we then get
\begin{equation}
e^{|U|\dt n_\up n_\dn}
=\frac{1}{2}\sum_{s=\pm 1}\prod_{\s=\up,\dn}
e^{\left(s\lambda + 
\frac{|U|\dt}{2}\right)\left(n_\s-\frac{1}{2}\right)},\ \ 
U<0,
\label{HSda}
\end{equation}
with $\lambda$ still being given by Eq.\ (\ref{lamb}).

The HS transformations then replace the on-site interaction by a
fluctuating field $s$ coupled to the magnetization or to the charge, in
the repulsive or attractive cases, respectively. As a consequence, the
argument of the exponential depends explicitly on $\s$ in the former case,
but not in the latter. As we will see below,
this very important difference is responsible for the absence of the
`minus-sign problem' in the attractive case.

We now replace the on-site interaction on every site of the space-time
lattice by Eqs.\ (\ref{HSdr}) or (\ref{HSda}), leading to the sought form 
in which only
bilinear terms appear in the exponential. For the repulsive case we get
\begin{widetext}
\begin{equation}
{\cal Z}=
\left(\frac{1}{2}\right)^{L^dM}
\underset{\{s\}}{\text{\Large\rm Tr}}\, 
\ctr
\prod_{\ell=M}^1 
\prod_{\s=\up,\dn}
e^{-\dt\sum_{i,j}c_{i\s}^{\dagger} K_{ij} 
c_{j\s}^{\phantom{\dagger}}}
e^{-\dt\sum_{i}c_{i\s}^{\dagger} V_{i}^\s(\ell) 
c_{i\s}^{\phantom{\dagger}}},
\label{Znew}
\end{equation}
\end{widetext}
where the traces are over Ising fields and over fermion 
occupancies on every site, and the product from $\ell=M$ to 1 simply
reflects the fact that earlier `times' appear to the right.
The time-slice index $\ell$ appears through the HS field $s_i(\ell)$ in   
\begin{equation}
V_i^\s(\ell)= \frac{1}{\dt} \lambda \s s_i(\ell)+ 
\left(\mu-\frac{U}{2}\right),
\label{Vi}
\end{equation}
which are the elements of the $N_s\times N_s$ diagonal matrix ${\sf 
V}^\s(\ell)$. 
One also needs the $N_s\times N_s$ hopping matrix ${\sf K}$, with elements
\begin{equation}
K_{ij}=
\begin{cases}
   -t & \text{if $i$ and $j$ are nearest neighbours},\\
   0  & \text{otherwise}
\end{cases}
\label{Kij}
\end{equation}
%are the matrix elements of the $N_s\times N_s$ hopping matrix ${\sf K}$. 
For instance, in one-dimension, and with periodic boundary conditions, one 
has an $L\times L$ matrix,
\begin{equation}
{\sf K} =\begin{pmatrix}
       {\ 0}&{-t}    &{\ 0}     &{\cdots}&{\ 0}     &{-t\ }\cr
        {-t}&{\ 0}   &{-t}      &{\cdots}&{\ 0}     &{\ 0}\cr   
       {\ 0}&{-t}    &{\ 0}     &{\cdots}&{\ 0}     &{\ 0}\cr   
    {\vdots}&{\vdots}&{\vdots}&{\vdots}&{\vdots}&{\vdots}\cr
        {-t}&{\ 0}   &{\cdots}&{\ 0}   &{-t}    &{\ 0}\cr
\end{pmatrix}
\label{k1}\ .
\end{equation} 

With bilinear forms in the exponential, the fermions can be traced out of
Eq.\ (\ref{Znew}) according to the development of Appendices 
\ref{group} and \ref{trace}.  
From Eq.\ (\ref{trace-det}), with the spin indices reintroduced, and with
the identification $e^{-\dt \mathsf{h}^\s(\ell)}\equiv
e^{-\dt{\mathsf{K}}} e^{-\dt{\mathsf{V}}^\s(\ell)}$, we have
\begin{equation}
{\mathcal Z}=\left(\frac{1}{2}\right)^{L^dM}
\underset{\{s\}}{\text{\Large\rm Tr}}\, \prod_{\s} 
\det \left[ {\mathbf{1}} + \bb_M^\s\bb_{M-1}^\s\ldots \bb_1^\s\right],
\label{detB}
\end{equation}
where we have defined
\begin{equation}
\bb_\ell^\s \equiv e^{-\dt{\mathsf{K}}} e^{-\dt{\mathsf{V}}^\s(\ell)},
\label{defB}
\end{equation}
in which the dependence with the auxiliary Ising spins has not been 
explicitly written, but should be understood, since they come in through 
the matrix ${\mathsf{V}}^\s(\ell)$. Introducing 
\begin{equation}
{\mathsf{O}}^\s(\{s\})\equiv 
{\mathsf{1}} + \bb_M^\s\bb_{M-1}^\s\ldots \bb_1^\s,
\label{O}
\end{equation}
we arrive, finally, at
\begin{eqnarray}
&{\mathcal{Z}}&=\left(\frac{1}{2}\right)^{L^dM}
\underset{\{s\}}{\text{\Large\rm Tr}}\,  
\det {\mathsf{O}}^\up(\{s\})\cdot \det {\mathsf{O}}^\dn(\{s\})=\nonumber\\
&&= \btr\ \rho(\{s\}),
\label{detO}
\end{eqnarray}
where the last equality defines an effective `density matrix', $\rho(\{s\})$.

We have then expressed the grand partition function as a sum over Ising
spins of a product of determinants. If the quantity under the Tr were
positive definite, it could be used as a Boltzmann weight to perform
importance sampling over the Ising configurations. However, the product of
determinants can indeed be negative for some configurations, giving rise
to the `minus-sign problem'; see Sec.\ \ref{minus}.

In order to implement the above framework, the need for
another approximation is evident from Eq.\ (\ref{defB}):
one needs to evaluate the exponential of the hopping matrix,
which, in the general case, is neither an analytically simple operation,
nor efficiently implemented numerically. By considering
again a one-dimensional system, we see that different powers of
${\mathsf{K}},$ as given by Eq.\ (\ref{k1}), generate many different
matrices. However, we can introduce a `checkerboard break-up' by writing
\begin{equation}
{\mathsf K}={\mathsf K}_x^{(a)} + {\mathsf K}_x^{(b)},
\label{cbx}
\end{equation}
such that ${\mathsf K}^{(a)}$ involves hoppings between sites 1 and 2, 3 and 4,$\ldots$, while ${\mathsf K}^{(b)}$ involves hoppings between 
sites 2 and 3, 4 and 5,$\ldots$. We now invoke the Suzuki-Trotter 
decomposition scheme to write
\begin{equation}
e^{-\dt{\mathsf{K}}}= e^{-\dt{\mathsf{K}}_x^{(a)}}e^{-\dt{\mathsf{K}}_x^{(b)}}
+{\cal O}\left[(\Delta\tau)^2\right],
\label{cbex}
\end{equation}
which leads to systematic errors of the same order as before. 
With this choice of break-up, even powers of 
${\mathsf {K}}_x^{(\alpha)},\ \alpha=a,b,$ 
become multiples of the identity matrix, and we end up with a
simple expression,
\begin{equation}
e^{-\dt{\mathsf{K}}_x^{(\alpha)}}= {\mathsf{K}}_x^{(\alpha)} \sinh \left(\dt t\right) +
{\mathsf{1}} \cosh \left(\dt t\right),
\label{coshdt}
\end{equation}
which is very convenient for numerical calculations. 
The breakup of Eq.\ (\ref{cbex}) can be generalized to three dimensions as 
follows
\begin{widetext}
\begin{equation}
e^{-\dt{\mathsf{K}}}=
e^{-\dt{\mathsf{K}}_z^{(a)}}e^{-\dt{\mathsf{K}}_y^{(a)}}e^{-\dt{\mathsf{K}}_x^{(a)}}
e^{-\dt{\mathsf{K}}_z^{(b)}}e^{-\dt{\mathsf{K}}_y^{(b)}}e^{-\dt{\mathsf{K}}_x^{(b)}}
+{\cal O}\left[(\Delta\tau)^2\right],
\label{K3d}
\end{equation}
\end{widetext}
where the separation along each cartesian direction is similar
to that for the one-dimensional case.

And, finally, we must discuss the calculation of average values.  For two
operators $A$ and $B$, their equal-`time' correlation function is
\begin{equation}
\ave{AB}=\frac{1}{\mathcal Z} \btr \ctr \left[ AB \prod_{\ell\s} 
e^{-\dt{\mathsf{K}}} e^{-\dt{\mathsf{V}}^\s(\ell)} \right].
\label{AB}
\end{equation}
If we now define the fermion average -- or Green's function -- 
for a given configuration of the HS 
fields as
\begin{equation}
\have{AB}\equiv 
\frac{1}{\rho(\{s\})}
\ctr \left[ AB \prod_{\ell\s} 
e^{-\dt{\mathsf{K}}} e^{-\dt{\mathsf{V}}^\s(\ell)} \right],
\label{gAB}
\end{equation}
the correlation function becomes
\begin{equation}
\ave{AB}= \frac{1}{\mathcal Z} \btr\  \have{AB}\, \rho(\{s\}).
\label{ABg}
\end{equation}

At this point it should be stressed the important role played by the Green's
functions in the simulations. Firstly, according to Eq.\ (\ref{ABg}), the
average value of an operator is straighforwardly obtained by sampling
the corresponding Green's function over the HS configurations weighted by
$\rho(\{s\})$. Secondly, as it will become apparent 
in Sec.\ \ref{qmc-s}, the single particle Green's 
function,  
$\have{c_{{\bf i}\sigma}^{\phantom{\dagger}}c_{{\bf 
j}\sigma}^{\dagger}}$,
plays a central role in the updating process itself. In 
Appendix \ref{etgf} we obtain
this quantity as the element ${\bf ij}$ of an $N_s\times
N_s$ matrix \cite{Hirsch85,vdl92}:
\begin{equation}
\have{c_{{\bf i}\sigma}^{\phantom{\dagger}}c_{{\bf 
j}\sigma}^{\dagger}}=
\left[\big({\mathsf{1}} + \bb_M^\s\bb_{M-1}^\s\ldots 
\bb_1^\s\big)^{-1}\right]_{{\bf i}{\bf j}},
\label{g1}
\end{equation}
which is, again, in a form suitable for numerical calculations. And,
thirdly, within the present approach the fermions only interact with the
auxiliary fields, so that Wick's theorem\cite{FW} holds \emph{for a fixed
HS configuration}\cite{Hirsch85,vdl92,Loh92}; the two-particle Green's
functions are then readily given in terms of the single-particle ones as
\begin{eqnarray}
&\have{c_{i_1}^{\dagger}c_{i_2}^{\phantom{\dagger}}c_{i_3}^{\dagger}
c_{i_4}^{\phantom{\dagger}}}&=
\have{c_{i_1}^{\dagger}c_{i_2}^{\phantom{\dagger}}}\, 
\have{c_{i_3}^{\dagger}c_{i_4}^{\phantom{\dagger}}}+\nonumber\\
&&+\ \have{c_{i_1}^{\dagger}c_{i_4}^{\phantom{\dagger}}}\, 
\have{c_{i_2}^{\phantom{\dagger}}c_{i_3}^{\dagger}},
\label{g2}
\end{eqnarray}
where the indices include spin, but since the $\up$ and $\dn$ spin
channels are factorized [\emph{c.f.} Eq.\ (\ref{gAB})], these fermion
averages are zero if the spins are different. All average values of
interest are therefore calculated in terms of single-particle Green's
functions.

As will be seen, unequal-time correlation functions are also important.
We define the operator $a$ in the `Heisenberg picture' as
\begin{equation}
a(\ell)\equiv a(\tau)=e^{\tau \mathcal{H}}\, a\, e^{-\tau \mathcal{H}},\ 
\tau\equiv \ell\dt, 
\label{heispic}
\end{equation}
so that the initial time is set to $\tau=\dt$ with this discretization,
and $a^\dagger(\ell)\neq [a(\ell)]^\dagger$. In Appendix \ref{utgf}, we
show that the unequal-time Green's function, for $\ell_1>\ell_2$, is 
given by \cite{Hirsch85}
\begin{eqnarray}
G_{{\bf i}{\bf j}}^\s(\ell_1;\ell_2)&\equiv&
\have{c_{{\bf i}\sigma}^{\phantom{\dagger}}(\ell_1)c_{{\bf 
j}\sigma}^{\dagger}(\ell_2)}=\nonumber\\
&=&\left[\bb_{\ell_1}^\s\bb_{\ell_1-1}^\s\ldots \bb_{\ell_2+1}^\s\,
\mathsf{g}^\s(\ell_2+1)\right]_{{\bf i}{\bf j}},
\label{gtau}
\end{eqnarray}
in which the Green's function matrix at the
$\ell$-th time slice is defined as
\begin{equation}
\mathsf{g}^\s(\ell)\equiv 
\left[ \mathsf{1} + 
\mathsf{A}^\s(\ell)\right]^{-1},
\label{gell}
\end{equation}
with
\begin{equation}
\mathsf{A}^\s(\ell)\equiv \bb_{\ell-1}^\s\bb_{\ell-2}^\s\ldots
\bb_{1}^\s\bb_{M}^\s\ldots\bb_{\ell}^\s.
\label{A}
\end{equation}

The reader should notice the order in which the products of 
$\mathsf{B}$'s are taken in Eqs.\ (\ref{g1}), (\ref{gtau}), 
and (\ref{A}); in Eq.\ 
(\ref{gtau}), in particular, the product runs from $\ell_2+1$ to $\ell_1$,
and not cyclically as in Eq.\ (\ref{A}).  Also, for a given configuration
$\{s\}$ of the HS spins, the equal-time Green's functions do display a
time-slice dependence, as expressed by Eq.\ (\ref{gell});  they only
become (approximately) equal after averaging over a large number of
configurations.

For future purposes, we also define
\begin{equation}
{\tilde g}_{\bf ij}^\s\equiv [\mathsf{1}-\mathsf{g}]_{\bf ij},
\ \ \text{and}\ \ 
{\tilde G}_{\bf ij}^\s\equiv [\mathsf{1}-\mathsf{G}]_{\bf ij}.
\end{equation}

\section{Determinantal Quantum Monte Carlo: The Simulations}
\label{qmc-s}

The simulations follow the lines of those for classical systems, except
for both the unusual Boltzmann weight, and the fact that one sweeps
through a space-time lattice. With the parameters of the Hamiltonian, $U$
and $\mu$, as well as the temperature, fixed from the outset, we begin by
generating, say a random configuration, $\{s\}$, for the HS fields. Since
the walker starts on the first time-slice, we use the definition, Eq.\
(\ref{g1}), to calculate the Green's function at $\ell=1$. As the walker
sweeps the spatial lattice, it attempts to flip the HS spin at every one
of the $N_s$ points.

At this point, it is convenient to picture the walker attempting to flip
the HS (Ising) spin on a generic site, $\lvec{i}$ of a generic time slice,
$\ell$.  If the spin is flipped, the matrices $\mathsf{B}_\ell^\up$ and
$\mathsf{B}_\ell^\dn$ change due to the element $\lvec{i}\lvec{i}$ of the
matrices $\mathsf{V}^\up(\ell)$ and $\mathsf{V}^\dn(\ell)$, respectively,
being affected; see Eqs.\ (\ref{Vi}) and (\ref{defB}). The expression for
the change in the matrix element, as $s_{\lvec{i}}(\ell)\to
-s_{\lvec{i}}(\ell)$, is
\begin{equation}
\delta V_{\lvec{i}\lvec{j}}^\s(\ell)\equiv 
V_{\lvec{i}\lvec{j}}^\s(\ell,-s)- V_{\lvec{i}\lvec{j}}^\s(\ell,s)
=-2\lambda\s s_{\lvec{i}}(\ell) \ \delta_{\lvec{i}\lvec{j}},
\label{deltaV}
\end{equation}
which allows us to write the 
change in $\mathsf{B}_\ell^\s$ as a matrix product,
\begin{equation}
\mathsf{B}_\ell^\s\to\left[\mathsf{B}_\ell^\s\right]^\prime= 
\mathsf{B}_\ell^\s\Delta_\ell^\s(\lvec{i}),
\label{deltaB}
\end{equation}
where the elements of the matrix $\mathsf{\Delta}_\ell^\s(\lvec{i})$ 
are
\begin{equation}
\left[\Delta_\ell^\s(\lvec{i})\right]_{\lvec{j}\lvec{k}}=
\begin{cases}
   0 & \text{if $\lvec{j}\neq\lvec{k}$},\\
   1 & \text{if $\lvec{j}=\lvec{k}\neq\lvec{i}$},\\
   e^{-2\lambda\s s_{\lvec{i}}(\ell)} & \text{if $\lvec{j}=\lvec{k}=\lvec{i}$}.
\end{cases}
\label{Deltaell}
\end{equation}

Let us now call $\{s\}^\prime$ and $\{s\}$, the HS 
configurations in which all Ising spins are the same,
except for those on site $(\lvec{i},\ell)$, which are opposite.
We can then write the ratio of `Boltzmann weights' as
\begin{equation}
r^\prime=\frac{\rho\left(\{s\}^\prime\right)}{\rho\left(\{s\}\right)}=
\frac{\det \mathsf{O}^\up\left(\{s\}^\prime\right)\cdot 
\det \mathsf{O}^\dn\left(\{s\}^\prime\right)}
{\det \mathsf{O}^\up\left(\{s\}\right)\cdot 
\det \mathsf{O}^\dn\left(\{s\}\right)}
=R_\up R_\dn,
\label{ratioprime}
\end{equation}
where we have defined the ratio of fermion determinants as
\begin{equation}
R_\s\equiv \frac{\det \mathsf{O}^\s\left(\{s\}^\prime\right)}
{\det \mathsf{O}^\s\left(\{s\}\right)}.
\label{Rs}
\end{equation}

It is important to notice that one actually does not need to calculate
determinants, since $R_\s$ is given in terms of the Green's 
function:
\begin{eqnarray}
&R_\s&=\frac{\det \left[\mathsf{1}+\mathsf{A}^\s(\ell)\mathsf{\Delta}_\ell^\s(\lvec{i})\right]}
{\det\left[ \mathsf{1}+\mathsf{A}^\s(\ell)\right]}=\nonumber\\
&&=
\det\left[ \big(\mathsf{1}+\mathsf{A}^\s(\ell)\mathsf{\Delta}_\ell^\s(\lvec{i})\big)
{\mathsf{g}^\s(\ell)}\right]=\nonumber\\
&&=\det \left[\mathsf{1}+\big(\mathsf{1}-\mathsf{g}^\s(\ell)\big)
\big(\mathsf{\Delta}_\ell^\s(\lvec{i})-\mathsf{1}\big)\right]=\nonumber\\
&&=1+\big(1-\mathsf{g}_{\lvec{i}\lvec{i}}^\s(\ell)\big)
\left(e^{-2\lambda\s s_{\lvec{i}}(\ell)}-1\right).
\label{Rsimple}
\end{eqnarray}
The last equality follows from the fact that 
\begin{equation}
\mathsf{\Gamma}_\ell^\s(\lvec{i})\equiv
\mathsf{\Delta}_\ell^\s(\lvec{i})-\mathsf{1}
\label{gamma}
\end{equation}
is a matrix such that all elements are zero, except for the 
$\lvec{i}$-th position
in the diagonal, which is $\gamma_\ell^\s(\lvec{i})\equiv e^{-2\lambda\s 
s_{\lvec{i}}(\ell)}-1$.
With this simple form for $R_\s$, the heat-bath algorithm is 
then easily implemented, with the probability of acceptance of
the new configuration being given by Eq.\ (\ref{heatbath}), with
Eqs.\ (\ref{ratioprime}) and (\ref{Rsimple}).

If the new configuration is accepted, the \emph{whole} Green's function
for the current time slice
must be updated, not just its element $\lvec{i}\lvec{i}$; this is the
non-local aspect of QMC simulations we referred to earlier.  There are two
ways of updating the Green's function. One can either compute the `new' 
one \emph{from scratch,} through Eq.\ (\ref{gell}), or \emph{iterate} the
`old' Green's function, by following along the lines that led to Eq.\
(\ref{Rsimple}), which yields
\begin{equation}
\bar{\mathsf{g}}^\s(\ell)=
\left[
  \mathsf{1}+ \big(\mathsf{1}-\mathsf{g}^\s(\ell)\big)
              \mathsf{\Gamma}_\ell^\s(\lvec{i})
\right]^{-1} 
\mathsf{g}^\s(\ell).
\label{gupdate}
\end{equation}
An explicit form for $\bar{\mathsf{g}}^\s(\ell)$ is obtained, e.g., by
first using an ansatz to calculate the inverse of the matrix in square
brackets above,
\begin{equation}
\left[
  \mathsf{1}+ \big(\mathsf{1}-\mathsf{g}^\s(\ell)\big)
              \mathsf{\Gamma}_\ell^\s(\lvec{i})
\right]^{-1}=
\mathsf{1}+ x\big(\mathsf{1}-\mathsf{g}^\s(\ell)\big)
              \mathsf{\Gamma}_\ell^\s(\lvec{i})
\label{ansatz}
\end{equation}
where $x$ is to be determined from the condition that the product of both 
matrices in Eq.\ (\ref{ansatz}) is $\mathsf{1}$. Using the fact that 
$\mathsf{\Gamma}_\ell^\s(\lvec{i})$ is very sparse, one arrives at
\begin{equation}
x= -\frac{1}{R_\s},
\label{xresult}
\end{equation}
with $R_\s$ being given by Eq.\ (\ref{Rsimple}). 
The element $\lvec{j}\lvec{k}$ of the Green's function is then 
updated according to
\begin{equation}
\bar{g}_{\lvec{j}\lvec{k}}^\s(\ell)=
g_{\lvec{j}\lvec{k}}^\s(\ell)-
\frac{\left[\delta_{\lvec{j}\lvec{i}}-g_{\lvec{j}\lvec{i}}^\s(\ell)\right]
\gamma_\ell^\s(\lvec{i})\,
g_{\lvec{i}\lvec{k}}^\s(\ell)}
{1+\left[1-g_{\lvec{i}\lvec{i}}^\s(\ell)\right]
\gamma_\ell^\s(\lvec{i})}.
\label{newg}
\end{equation}
Alternatively, one could arrive at the same result by solving a Dyson's 
equation for $\bar{g}_{\lvec{j}\lvec{k}}^\s(\ell)$ \cite{vdl92}.

After the walker tries to flip the spin on the last site of the $\ell$-th
time slice, it moves on to the first site of the ($\ell+1$)-th time slice.  
We therefore need the Green's function for the ($\ell+1$)-th time slice,
which, as before, can be calculated either from scratch, or iteratively
from the Green's function for the $\ell$-th time slice. Indeed, by
comparing $\left[\mathsf{g}^\s(\ell+1)\right]^{-1}$ with
$\left[\mathsf{g}^\s(\ell)\right]^{-1}$, as given by Eq.\ (\ref{gell}), it
is easy to see that
\begin{equation}
\mathsf{g}^\s(\ell+1)=\bb_\ell^\s \mathsf{g}^\s(\ell) 
\big[\bb_\ell^\s\big]^{-1},
\label{gell1}
\end{equation}
which can be used to compute the Green's function in the subsequent time 
slice.

While the computation from scratch of the new Green's function takes $\sim
N_s^3$ operations, the iterations [Eq.\ (\ref{newg}) and (\ref{gell1})]
only take $\sim N_s^2$ operations. Hence, at first sight the latter form
of updating should be used. However, rounding errors will be amplified
after many iterations, thus deteriorating the Green's function calculated
in this way; see Sec.\ \ref{stable}. A solution of compromise is to iterate the Green's function
while the walker sweeps all spatial sites of $\tilde{\ell}$ ($\sim 10$)
time slices, and then to compute a new one from scratch when the
($\tilde{\ell}+1$)-th time slice is reached. This `refreshed' Green's
function will then be used to start the iterations again. Clearly, one
should check $\mathsf{g}$ for accuracy, by comparing the updated and
refreshed ones at the ($\tilde{\ell}+1$)-th time slice; if the accuracy is
poor, $\tilde{\ell}$ must be decreased.

For completeness, we now return to the situation described in the
beginning of this Section, in which the walker attempts to flip the HS
spins on every spatial site of the $\ell=1$ time slice. Whenever the flip
is accepted, the Green's function is updated according to Eq.\
(\ref{newg}). After sweeping all spatial sites, and before the walker
moves on to the first site of the $\ell=2$ time slice, we use Eq.\
(\ref{gell1}) to calculate $\mathsf{g}^\s(2)$. The walker then sweeps the
spatial sites of this time-slice, attempting to flip every spin; if the
flip is accepted, the Green's function is updated according to Eq.\
(\ref{newg}). As mentioned above, this procedure is repeated for the next
$\tilde{\ell}$ time slices. After sweeping the last one of these, a new
Green's function is calculated from the definition, Eq.\ (\ref{gell}).
For the subsequent $\tilde{\ell}$ time slices the iterative computation of 
$\mathsf{g}$ is used, and so forth.

Similarly to the classical case, one sweeps the space-time lattice several
times (warm-up sweeps) before calculating average values. After warming
up, one starts the `measuring' stage, in which another advantage of the
present approach manifests itself in the fact, already noted, that all
average quantities are expressed in terms of the Green's functions. This
brings us to the discussion of the main quantities used to probe the
physical properties of the system, and how they relate to the Green's
functions.

We start with the calculation of the occupation, $n$. It is given by
\begin{eqnarray}
&n&=\frac{1}{MN_s}\sum_{\lvec{i},\ell}\sum_{\s}
\ave{
      c_{\lvec{i}\s}^\dagger(\ell)
      c_{\lvec{i}\s}^{\phantom{\dagger}}(\ell)
    }=
\nonumber\\
&&=
\frac{1}{MN_s}\sum_{\lvec{i},\ell}\sum_\s
\dave{1-g_{\lvec{i}\lvec{i}}^\s(\ell)},
\label{occn}
\end{eqnarray}
where the first equality illustrates that the ensemble average ({\it
i.e.,} over both fermionic \emph{and} HS fields)
$\ave{n_{\lvec{i}\up}(\ell)+n_{\lvec{i}\dn}(\ell)}$ is itself averaged
over all $MN_s$ sites of the space-time lattice. In the second equality
the fermion degrees of freedom have been integrated out (hence the Green's
functions), and the average over HS fields, represented by double
brackets, is to be performed by importance sampling over $N_a$ HS
configurations [\emph{c.f.} Eq.\ (\ref{ABg})]. We then have
\begin{equation}
\dave{g_{\lvec{i}\lvec{j}}^\s(\ell)}\simeq\frac{1}{N_a}\sum_{\zeta=1}^{N_a}
g_{\lvec{i}\lvec{j}}^\s(\ell). 
\label{dave} 
\end{equation}

It should be reminded that since these are grand-canonical simulations,
the chemical potential must be chosen \emph{a priori} in order to yield
the desired occupation. The chemical potential leading to half filling in
the Hubbard model (with nearest-neighbour hopping only) is obtained by
imposing particle-hole symmetry \cite{Emery76,dS93}: one gets $n=1$ for
$\mu=U/2$ for all $T$ and $N_s$. Away from half-filling, one has to
determine $\mu$ by trial and error, which must be done repeatedly, since
the dependence of $n$ with $\mu$ (for given $N_s$ and $U$) itself varies
with temperature. Overall, the dependence of $n$ with $\mu$ indicates
that the Hubbard model is an insulator at half filling in any dimension:
indeed, the compressibility
\begin{equation}
\kappa\equiv -\frac{1}{V}\left(\frac{\partial V}{\partial P}\right)_{T,\mu}=
\frac{1}{n^2}\left(\frac{\partial n}{\partial \mu}\right)_{T},
\label{compress}
\end{equation}
where $V$ and $P$ are the volume and pressure, respectively, vanishes
at $\mu=U/2$, so that no particles can be added to the system; see, 
e.g., Ref.\ \cite{Moreo90} for explicit data for $n(\mu)$ in two 
dimensions.

The magnetic properties are probed in several ways. 
Since the components of the magnetization operator are
\begin{subequations}
\label{mags}
\begin{eqnarray}
m_{\lvec{i}}^x&\equiv&
c_{\lvec{i}\up}^\dagger c_{\lvec{i}\dn}^{\phantom{\dagger}}+ c_{\lvec{i}\dn}^\dagger c_{\lvec{i}\up}^{\phantom{\dagger}},
\label{mx}
\\
m_{\lvec{i}}^y&\equiv&
-i\left(c_{\lvec{i}\up}^\dagger c_{\lvec{i}\dn}^{\phantom{\dagger}}- c_{\lvec{i}\dn}^\dagger c_{\lvec{i}\up}^{\phantom{\dagger}}\right),
\label{my}
\\
m_{\lvec{i}}^z&\equiv&n_{\lvec{i}\up}-n_{\lvec{i}\dn},
\label{mz}
\end{eqnarray}
\end{subequations}
Wick's theorem [Eq.\ (\ref{g2})] allows one to obtain 
the equal-time spin-density correlation 
functions as
\begin{subequations}
\label{sdw}
\begin{eqnarray}
\ave{m_{\lvec{i}}^zm_{\lvec{j}}^z}&=&
\dave{\,
  \gt_{\lvec{i}\lvec{i}}^\up\,\gt_{\lvec{j}\lvec{j}}^\up+
  \gt_{\lvec{i}\lvec{j}}^\up\,g_{\lvec{i}\lvec{j}}^\up -
  \gt_{\lvec{i}\lvec{i}}^\up\,\gt_{\lvec{j}\lvec{j}}^\dn-\nonumber\\
&& 
  \quad 
   -\gt_{\lvec{i}\lvec{i}}^\dn\,\gt_{\lvec{j}\lvec{j}}^\up+
   \gt_{\lvec{i}\lvec{i}}^\dn\,\gt_{\lvec{j}\lvec{j}}^\dn+
   \gt_{\lvec{i}\lvec{j}}^\dn\,g_{\lvec{i}\lvec{j}}^\dn \,
},
\label{sdwz}
\\
\ave{m_{\lvec{i}}^xm_{\lvec{j}}^x}&=&
\dave{\,
  \gt_{\lvec{i}\lvec{j}}^\up\,g_{\lvec{i}\lvec{j}}^\dn+
  \gt_{\lvec{i}\lvec{j}}^\dn\,g_{\lvec{i}\lvec{j}}^\up
},
\label{sdwx}
\end{eqnarray}
\end{subequations}
and a similar expression for the $yy$ correlations. It is
understood that the $g$'s are to be evaluated at the 
same time slice as the $m$'s on the LHS.

If spin rotational symmetry is preserved, as in the case of
a singlet ground state, Eqs.\ (\ref{sdwz}) and (\ref{sdwx})
should yield the same result. This is indeed the case for 
average values, but,
in practice, transverse ({\it i.e.,} $xx$ and $yy$) 
correlations are much less noisy than the longitudinal ones
\cite{Hirsch3D}; this can be traced back to the discrete HS 
transformation, which singles out the $z$ component by coupling 
the auxiliary fields to the $m^z$'s. It should also be pointed
out that a possible ferromagnetic state (either saturated or 
just a non-singlet one) would break this symmetry. 

Setting $\lvec{i}=\lvec{j}$ in Eqs.\ (\ref{sdw}) leads to the
local moment,
\begin{equation}
\ave{m_\lvec{i}^2}=\bave{\left(m_\lvec{i}^x\right)^2+
\left(m_\lvec{i}^y\right)^2+\left(m_\lvec{i}^z\right)^2}
%\dave{\,\tilde g_\lvec{ii}^\up+\tilde g_\lvec{ii}^\dn
%-2\,\tilde g_\lvec{ii}^\up\,\tilde g_\lvec{ii}^\dn},
\label{local-mom}
\end{equation}
which measures the degree of itinerancy of the fermions. 
For instance, considering again the Hubbard model at half filling,  
$\ave{\left(m_\lvec{i}^\nu\right)^2},\ \nu=x,y,z,$ decreases
from 1, in the frozen-charge state ({\it i.e.,} when $U\to\infty$),
to 1/2, in the metallic state ($U=0$). 

One can collect the contributions to the spin-spin correlation 
functions from different sites, and calculate the equal-time
magnetic structure factor as
\begin{equation}
S(\mathbf{q})=\frac{1}{N_s}\sum_{\lvec{i}\lvec{j}} 
e^{i\mathbf{q}\cdot (\lvec{i}-\lvec{j})}
\ave{\mathbf{S}_{\lvec{i}}\cdot\mathbf{S}_{\lvec{j}}},
\label{sofq}
\end{equation}
where $S_{\lvec{i}}^\nu=m_{\lvec{i}}^\nu/2$, since $\hbar=1$.
$S(\mathbf{q})$ shows peaks at values of $\mathbf{q}$ 
corresponding to the dominant magnetic arrangements. 
For the Hubbard model in one dimension, $S(q)$ has peaks at
$q=2k_F=\pi n$, corresponding to quasi-long range order 
(\emph{i.e.,} algebraic spatial decay of correlations) in the
ground state; one refers to this as a state with a 
\emph{spin-density wave}\cite{Gruner1}. In two dimensions, 
and at half filling, the peak
located at $\mathbf{q}=(\pi,\pi)$ \cite{HirschTang89} signals
N\'eel order in the ground state; away from half-filling, the
system is a paramagnet \cite{Hirsch85,HirschTang89} with strong 
AFM correlations at incommensurate $\mathbf{q}$'s \cite{Moreo90}.
 
The non-commutation aspects imply that the susceptibility is 
given by \cite{FW}
\begin{equation}
\chi(\mathbf{q})=\frac{1}{N_s}\sum_{\lvec{i}\lvec{j}}
e^{i\mathbf{q}\cdot (\lvec{i}-\lvec{j})}
\int_0^\beta d\tau\ \ave{\mathbf{S}_{\lvec{i}}(\tau)\cdot\mathbf{S}_{\lvec{j}}}.
\label{magsusc}
\end{equation}
In the discrete-time formulation, the integral becomes a sum over 
different time intervals, which is carried out with the aid of the
unequal-time Green's functions. Focusing on one of the components
of the scalar product above, say $z$, we have
\begin{eqnarray}
&\ave{m_{\lvec{i}}(\ell_1)m_{\lvec{j}}(\ell_2)}=
&\sum_{\s,\s^\prime} 
\Big[\left(2\delta_{\s\s^\prime}-1\right)
\bigdave{
\tilde g_{\lvec{i}\lvec{i}}^\s(\ell_1)
 \tilde g_{\lvec{j}\lvec{j}}^{\s^\prime}(\ell_2)}+\nonumber\\
&&+\delta_{\s\s^\prime}
  \bigdave{\tilde G_{\lvec{i}\lvec{j}}^\s(\ell_1;\ell_2)
  G_{\lvec{i}\lvec{j}}^\s(\ell_1;\ell_2)}
\Big].
\label{sdwt}
\end{eqnarray}
  
Similarly to the magnetic properties, one can also investigate whether 
or not there is the formation of a charge density wave \cite{Gruner2,Gruner3}.
With $n_{\lvec{i}}=n_{\lvec{i}\up}+n_{\lvec{i}\dn}$, the \emph{equal-time}
correlation function is given by
\begin{equation}
\ave{n_{\lvec{i}}n_{\lvec{j}}}=
\sum_{\s,\s^\prime} 
\Big[
\bigdave{
\tilde g_{\lvec{i}\lvec{i}}^\s\,
 \tilde g_{\lvec{j}\lvec{j}}^{\s^\prime}}
+\delta_{\s\s^\prime}
  \bigdave{\tilde g_{\lvec{j}\lvec{i}}^\s\,
  g_{\lvec{i}\lvec{j}}^\s}
\Big],
\label{cdw}
\end{equation}
in terms of which the charge structure factor becomes
\begin{equation}
\mathcal{C}(\mathbf{q})=\frac{1}{N_s}\sum_{\lvec{i}\lvec{j}} 
e^{i\mathbf{q}\cdot (\lvec{i}-\lvec{j})}
\ave{n_{\lvec{i}}n_{\lvec{j}}}.
\label{cofq}
\end{equation}
It is also useful to define a charge susceptibility as
\begin{equation}
\mathcal{N}(\mathbf{q})=\frac{1}{N_s}\sum_{\lvec{i}\lvec{j}} 
e^{i\mathbf{q}\cdot (\lvec{i}-\lvec{j})}
\int_0^\beta d\tau\ \ave{n_{\lvec{i}}(\tau)n_{\lvec{j}}},
\label{ch-susc}
\end{equation}
where, in the discrete time formulation, the above average values 
are given in terms of the unequal-time Green's functions as
\begin{eqnarray}
&\ave{n_{\lvec{i}}(\ell_1)n_{\lvec{j}}(\ell_2)}=
&\sum_{\s,\s^\prime} 
\Big[
\bigdave{
\tilde g_{\lvec{i}\lvec{i}}^\s(\ell_1)
 \tilde g_{\lvec{j}\lvec{j}}^{\s^\prime}(\ell_2)}+\qquad\qquad
\nonumber\\
&&+\delta_{\s\s^\prime}
  \bigdave{\tilde G_{\lvec{j}\lvec{i}}^\s(\ell_1;\ell_2)
  G_{\lvec{i}\lvec{j}}^\s(\ell_1;\ell_2)}
\Big];
\label{cdwt}
\end{eqnarray}
it should be noted that the uniform charge susceptibility
is proportional to the compressibility, defined by Eq.\ 
(\ref{compress}).

For the one-dimensional Hubbard model, the presence of a 
charge-density wave (CDW) is signalled by a cusp in 
$\mathcal{C}(\mathbf{q})$ and by a divergence (as $T\to 0$)
in $\mathcal{N}(\mathbf{q})$. While the insulating character
at half filling prevents a CDW from setting in, 
away from half filling there has been some debate on whether the 
position of the cusp is at $q=2k_F$, as predicted by
the Luttinger liquid theory\cite{Schulz90}, or at $q=4k_F$, as 
evidenced by QMC data and Lanczos diagonalizations \cite{tclp00}.
In two dimensions, and away from half filling, the situation is 
even less clear.

We now discuss superconducting correlations. The imaginary-time
singlet pair-field operator is a generalization of the BCS gap
function \cite{Tilley},
\begin{equation}
\Delta_\zeta(\tau)\equiv\frac{1}{\sqrt{N_s}}\sum_{\mathbf{k}} f_\zeta(\mathbf{k})
c_{\mathbf{k}\up}^{\phantom{\dagger}}(\tau)
c_{-\mathbf{k}\dn}^{\phantom{\dagger}}(\tau),
\label{Delta_r}
\end{equation}
where the subscript $\zeta$ in the form factor, 
$f_\zeta(\mathbf{k})$, labels the symmetry of the
pairing state, following closely the notation for hydrogenic
orbitals; for instance, in two dimensions one has
\begin{subequations}
\label{f_r}
\begin{eqnarray}
\text{$s$-wave:}\ f_s(\mathbf{k})=& 1,
\label{s-wave}
\\
\text{extended $s$-wave:}\ f_{s^*}(\mathbf{k})=& \cos k_x + \cos k_y,
\label{s*-wave}
\\
\text{$d_{x^2-y^2}$-wave:}\ f_{d_{x^2-y^2}}(\mathbf{k})=& \cos k_x - \cos k_y,
\label{d-wave}
\\
\text{$d_{xy}$-wave:}\ f_{d_{xy}}(\mathbf{k})=& \sin k_x  \sin k_y,
\label{dxy-wave}
\end{eqnarray}
\end{subequations}
and several other possibilities, including triplet pairing, as well 
as linear combinations of the above. Ignoring for the time being 
any $\tau$-dependence, and introducing the Fourier transform 
of the annihilation operators, Eq.\ (\ref{Delta_r}) can be written 
as
\begin{equation}
\Delta_\zeta=\frac{1}{\sqrt{N_s}}\sum_{\lvec{i}}\Delta_\zeta(\lvec{i})
\label{Delta_real}
\end{equation}
with the site-dependent pair-field operator being given by
\begin{equation}
\Delta_\zeta(\lvec{i})=\frac{1}{2}\sum_{\mathbf{a}}\tilde{f}_\zeta ({\mathbf{a}})
c_{\lvec{i}\up}c_{\lvec{i}+{\mathbf{a}}\dn},
\label{Delta_i}
\end{equation}
where the sum runs over lattice sites neighbouring 
$\lvec{i}$, with range and relative phase determined by 
\begin{equation}
\tilde{f}_\zeta({\mathbf{a}})=
\frac{2}{N_s}\sum_{\mathbf{k}}f_\zeta(\mathbf{k})e^{-i\mathbf{k}\cdot\mathbf{a}}.
\label{ftilde}
\end{equation} 
For the $d_{x^2-y^2}$-wave, for instance, 
\begin{equation}
\tilde{f}_{d_{x^2-y^2}}({\mathbf{a}})=\frac{2}{N_s}\left(\delta_{\mathbf{a},\mathbf{x}}+
\delta_{\mathbf{a},-\mathbf{x}}-
\delta_{\mathbf{a},\mathbf{y}}-
\delta_{\mathbf{a},-\mathbf{y}}\right),
\label{fd}
\end{equation}
where $\mathbf{x}$ and $\mathbf{y}$ are unit vectors along the cartesian
directions.

Similarly to the magnetization, simple averages of pair-field
operators vanish identically in a finite-sized system, so one again
turns to correlation functions. 
We define the equal-time \emph{pairing correlation function} as
\begin{equation}
P_\zeta (\lvec{i},\lvec{j})=
\ave{\Delta_\zeta^\dagger(\lvec{i}) 
\Delta_\zeta^{\phantom{\dagger}}(\lvec{j})+{\text{H.c.}}},
\label{Pzeta}
\end{equation}
whose spatial decay is sometimes analysed as a probe for the 
superconducting state \cite{Kuroki}. Its uniform ({\it i.e.,} 
$\mathbf{q}=0$) Fourier transform,
\begin{equation}
\tilde{P}_\zeta = \frac{1}{N_s}\sum_{\lvec{i},\lvec{j}}
\ave{\Delta_\zeta^\dagger(\lvec{i}) 
\Delta_\zeta^{\phantom{\dagger}}(\lvec{j})+{\text{H.c.}}},
\end{equation}
has also been used, together with finite-size scaling
arguments, to probe superconducting pairing 
\cite{Moreo91a,dS93,dS94}. As before, it is a straightfoward 
exercise to express the above averages in terms of HS averages 
of Green's functions.

Restoring the $\tau$-dependence, another
useful probe is the uniform and
zero-frequency ($\omega=0$) \emph{pairing susceptibility}, 
\begin{equation}
\Pi_\zeta=\frac{1}{N_s}\sum_{\lvec{i},\lvec{j}}\int_0^\beta
d\tau\ 
\ave{\Delta_\zeta^\dagger(\lvec{i},\tau) 
\Delta_\zeta^{\phantom{\dagger}}(\lvec{j})+{\text{H.c.}}},
\label{Pi}
\end{equation}
A considerable enhancement of $\Pi_\zeta$ over its non-interacting 
counterpart may be taken as indicative
of a superconducting state; see, {\it e.g.,} Refs.\ \cite{HirschLin88}
and \cite{dS89}. In fairness, there has been no real consensus 
on which is the best numerical probe of a superconducting state.
It may be argued, for instance, that $P_\zeta$ should be compared
with the corresponding nonvertex correlation function; see, e.g.,
Refs.\ \cite{White89,Moreo91}. 

While the above probes for superconductivity assume a given symmetry
for the pairing state, there is an alternative procedure which 
is free from such assumption \cite{SWZ}. The Kubo linear response
to a vector potential $A_x(\mathbf{q},\omega)$ is given by the 
$x$-component of the current density,
\begin{equation}
\ave{j_x(\mathbf{q},\omega)}= -e^2\left[\ave{-K_x}-\Lambda_{xx}(\mathbf{q},\omega)\right]A_x(\mathbf{q},\omega),
\label{Kubo}
\end{equation}
where
\begin{equation}
\ave{K_x}=\Bave{-t \sum_\sigma
            \left(c_{\lvec{i}+\mathbf{x}\sigma}^\dagger 
                   c_{\lvec{i}\sigma}^{\phantom{\dagger}}+ 
                  c_{\lvec{i}\sigma}^\dagger 
                   c_{\lvec{i}+\mathbf{x}\sigma}^{\phantom{\dagger}}
            \right)}
\label{Kx}
\end{equation}
is the kinetic energy associated with the $x$-oriented links, and
\begin{equation}
\Lambda_{xx} (\mathbf{q}, i\omega_m) = 
\sum_{\lvec{i}}\int_0^\beta d\tau\ \ave{j_x(\lvec{i},\tau)j_x(0,0)}
e^{i\mathbf{q}\cdot\lvec{i}}e^{-i\omega_m\tau}
\label{Lambdaqomega}
\end{equation}
is the imaginary-time and space Fourier transform 
($\omega_m=2\pi m\beta$ is the Matsubara frequency \cite{FW})
of the \emph{current density correlation function,}
\begin{equation}
\Lambda_{xx} (\lvec{i}, \tau) \equiv \ave{j_x(\lvec{i},\tau)j_x(0,0)},
\label{Lambdaitau}
\end{equation}
with
\begin{equation}
j_x(\lvec{i},\tau)=e^{\tau\mathcal{H}}
  \left[
        it\sum_\sigma
            \left(c_{\lvec{i}+\mathbf{x}\sigma}^\dagger 
                  c_{\lvec{i}\sigma}^{\phantom{\dagger}}
                  - 
                  c_{\lvec{i}\sigma}^\dagger  
                  c_{\lvec{i}+\mathbf{x}\sigma}^{\phantom{\dagger}}
            \right)
  \right]
e^{-\tau\mathcal{H}}.
\label{jx}
\end{equation}
As an exercise, the reader should express the current density 
correlation function in terms of HS averages of Green's
functions.

The superconducting properties of the system are associated with
its long wavelength static response ({\it i.e.,} $\mathbf{q}\to 0, \omega=0$), but with a careful distinction in the way in which the
order of the limits $q_x\to 0$ (Longitudinal) and $q_y\to 0$ 
(Transverse) are taken: 
\begin{equation}
\Lambda^L\equiv \lim_{q_x\to 0} \Lambda_{xx}(q_x,q_y=0,i\omega_m=0),
\label{LambdaL}
\end{equation}
and
\begin{equation}
\Lambda^T\equiv \lim_{q_y\to 0} \Lambda_{xx}(q_x=0,q_y,i\omega_m=0).
\label{LambdaT}
\end{equation}
As a result of gauge invariance \cite{SWZ}, one should always have 
\begin{equation}
\Lambda^L=\ave{-K_x},
\label{gauge_inv}
\end{equation}
which can be used to check the numerics. On the other hand, the 
superfluid weight, $D_s$, (which is proportional to the superfluid 
density, $\rho_s$) turns out to be \cite{SWZ}
\begin{equation}
\frac{D_s}{\pi e^2}=\rho_s=\ave{-K_x}-\Lambda^T,
\label{D_s}
\end{equation}
so that a Meissner state is associated with $\Lambda^L\neq\Lambda^T$.
In practice, $\ave{-K_x}$ is calculated independently from 
$\Lambda^T$, and one checks whether the latter quantity 
approaches the former as $q_y\to 0$. Since the calculations
are performed on finite-sized systems, one should examine 
the trend of the data as the number of sites is increased; 
see Ref.\ \cite{SWZ}.

If the system is not a superconductor, the current-density
correlation function still 
allows us to distinguish between a metallic and an insulating 
state. Indeed, the limit $\mathbf{q}=0,\ \omega\to 0$ (note
the order of the limits, which is opposite to that used in
relation to the superfluid density!) of the conductivity 
determines whether or not the ground state of the system
has zero resistance \cite{SWZ}. The real part of the 
zero temperature conductivity can be written in the form
$\sigma_{xx}=D\delta(\omega)+\sigma_{xx}^{\mathrm{reg}}(\omega)$,
where $\sigma_{xx}^{\mathrm{reg}}(\omega)$ is a regular function
of $\omega$, and the Drude weight, $D$, describes the DC response.
The latter can be estimated at low temperatures by \cite{SWZ}
\begin{equation}
\frac{D}{\pi e^2}\simeq 
\left[ 
      \ave{-K_x}-\Lambda_{xx}(\mathbf{q}=0,i\omega_m\to 0)
\right],
\label{Drude}
\end{equation}
which amounts to extrapolating discrete non-zero $\omega_m$ 
data towards $\omega_m=0$. This can lead to a subtle interplay
between the energy scales involved; see Ref.\ \cite{SWZ} for
examples.

The above-mentioned criteria to determine whether the system
is metallic, insulating, or superconducting are summarized in 
Table \ref{table1}. The reader is urged to consult References
\cite{SWZ,Trivedi96,Scalettar99} for illustrative discussions 
on the many aspects involved in the data analyses.

\begin{table}
\caption{\label{table1} Criteria to determine whether a system 
is a superconductor, a metal, or an insulator, from the behaviour 
of the current-current correlation function.}
\begin{ruledtabular}
\begin{tabular}{lcr}
Nature of the state & $D_s$ [Eq.\ (\ref{D_s})] & $D$ [Eq.\ (\ref{Drude})]\\
\hline
Superconducting & $\neq 0$ & $\neq 0$\\
Metallic & 0 & $\neq 0$\\
Insulating & 0 & 0\\
\end{tabular}
\end{ruledtabular}
\end{table}

At this point one should already be convinced that a very wide
range of probes is available to scrutinize most of the possible
physical properties of a model system. However, two technical 
problems must be dealt with, which will be analysed in turn:
the minus-sign of the fermionic determinant, and the instabilities
appearing at low temperatures.

\section{The Minus-Sign Problem}
\label{minus}

As mentioned before, the product of fermionic determinants is not 
positive definite, apart from in a few cases. The best known instance
is the repulsive Hubbard model at half filling. To see why this is so, 
we employ a particle-hole transformation, 
\begin{equation}
c_{\lvec{i}\sigma}^\dagger\to d_{\lvec{i}\sigma}^{\phantom{\dagger}}=
(-1)^{\lvec{i}} c_{\lvec{i}\sigma}^\dagger,\ \
c_{\lvec{i}\sigma}^{\phantom{\dagger}}\to d_{\lvec{i}\sigma}^\dagger=
(-1)^{\lvec{i}} c_{\lvec{i}\sigma}^{\phantom{\dagger}}
\label{pht1}
\end{equation}
such that
\begin{equation}
n_{\lvec{i}\sigma}\equiv
c_{\lvec{i}\sigma}^\dagger c_{\lvec{i}\sigma}^{\phantom{\dagger}}
=1-d_{\lvec{i}\sigma}^\dagger d_{\lvec{i}\sigma}^{\phantom{\dagger}}
\equiv 1-\tilde n_{\lvec{i}\sigma}
\label{pht2}
\end{equation}
and consider the fermionic determinant at the symmetric point, 
$\mu=U/2$. One then has
\begin{eqnarray}
\det \mathsf{O}^\up &= &
  \underset{\{n\}} \ctr \prod_\ell e^{-\dt \mathsf{K}_\up} 
   e^{\lambda\sum_i s_i(\ell)n_{\lvec{i}\up}}= \nonumber\\
  &= &\underset{\{\tilde n\}} \ctr \prod_\ell e^{-\dt 
\tilde{\mathsf{K}}_\up} 
   e^{-\lambda\sum_i s_i(\ell)\tilde n_{\lvec{i}\up}}e^{-\lambda\sum_i s_i(\ell)}= \nonumber\\
&= & e^{-\lambda\sum_{i\ell} s_i(\ell)}\det \mathsf{O}^\dn, 
\label{phs-det}
\end{eqnarray}
where the tildes denote hole variables. Therefore,
\begin{equation}
\det \mathsf{O}^\up \cdot \det \mathsf{O}^\dn > 0,\ \ \text{for}\ n=1,\ U\geq 0. 
\end{equation}

\begin{figure} 
{\centering\resizebox*{3.4in}{!}{\includegraphics*{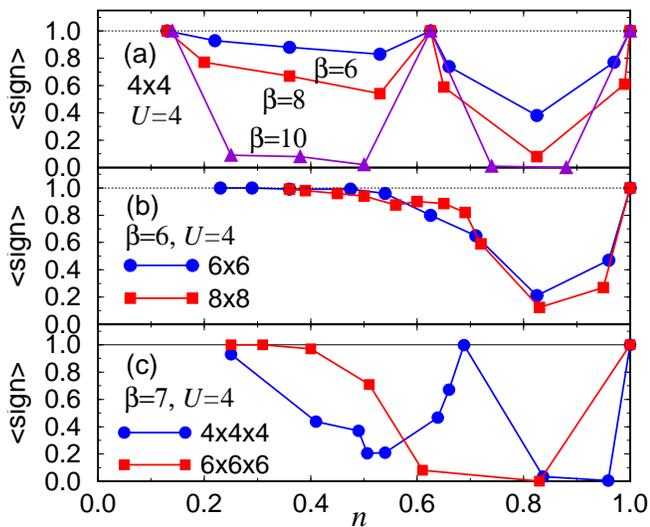}}}
\caption{\label{fig1} The average sign of the product of fermionic
determinants as a function of band filling, for the Hubbard model with
$U=4$: (a) $4\times 4$ square lattice, for inverse temperatures $\beta =
6$ (circles), 8 (squares), and 10 (triangles); adapted from Refs.\
\cite{vdl92} and \cite{White89}. (b) $6\times6$ (circles) and $8\times8$
(squares) square lattice, for fixed inverse temperature, $\beta=6$;
adapted from Ref.\ \cite{White89}.
(c) $4\times4\times4$ (circles) and $6\times6\times6$ (squares)
simple cubic lattice, for 
fixed  inverse temperature, $\beta=7$. Lines are guides to the eye in all 
cases.  
} 
\end{figure}

For the  \emph{attractive} Hubbard model, the lack of $\s$-dependence 
in the discrete HS transformation [see the comments below Eq.\ 
(\ref{HSda})] leads to
${\mathsf{O}}^\up(\{s\})\equiv{\mathsf{O}}^\dn(\{s\})$, 
so that the product of determinants is positive \emph{for all fillings.} 
Similar arguments apply to show that the fermionic determinant is 
always positive for the Holstein model for electron-phonon 
interactions \cite{Loh90,Noack91}.

In other cases, the fermionic determinant becomes negative for some
configurations. In order to circumvent this problem, recall that the
partition function can be written as a sum over configurations,
$c\equiv\{s\}$, of the `Boltzmann weight', $p(c)\equiv \det
\mathsf{O}^\up(\{s\}) \det\mathsf{O}^\dn(\{s\})$. If we write
$p(c)=s(c)|p(c)|$, where $s(c)=\pm1$ to keep track of the sign of $p(c)$,
the average of a quantitity $A$ can be replaced by an average weighted by
$|p(c)|$ as follows
\begin{eqnarray}
\ave{A}_p&=&\frac{\sum_c p(c)A(c)}{\sum_c p(c)}=
\frac{\sum_c |p(c)|s(c)A(c)}{\sum_c |p(c)|s(c)}=\nonumber\\
&=&\frac{\left[\sum_c |p(c)|s(c)A(c)\right]/\sum_c |p(c)|}
{\left[\sum_c |p(c)|s(c)\right]/\sum_c |p(c)|}=\nonumber\\
&=&\frac{\sum_c p^\prime(c)\left[s(c)A(c)\right]}{\sum_c p^\prime(c)\left[s(c)\right]}
\equiv \frac{\ave{sA}_{p^\prime}}{\ave{s}_{p^\prime}},
\label{pp_ave}
\end{eqnarray}
where $p^\prime(c)\equiv |p(c)|$. Therefore, if the absolute value of
$p(c)$ is used as the Boltzmann weight, one pays the price of having to
divide the averages by the \emph{average sign of the product of fermionic
determinants}, $\ave{sign}\equiv\ave{s}_{p^\prime}$. Whenever this
quantity is small, much longer runs (in comparison to the cases in 
which $\ave{sign}\simeq1$) are necessary to compensate the strong 
fluctuations in $\ave{A}_p$. Indeed, from Eq.\ (\ref{error}) we can 
estimate that the runs need to be stretched by a factor on the order of 
$\ave{sign}^{-2}$ in order to obtain the same quality of data as for 
$\ave{sign}\simeq 1$. 

In Fig.\ \ref{fig1}(a), we show the behaviour of $\ave{sign}$ as a
function of band filling, for the Hubbard model on a $4\times4$ square
lattice with $U=4$, and for three different temperatures. One sees that,
away from $n=1$, $\ave{sign}$ is only well behaved at certain fillings,
corresponding to `closed-shell' configurations; i.e., those such that the
ground state of the otherwise free system is non-degenerate
\cite{White89}. For the case at hand, the special fillings are 2 and 10
fermions on $4\times4$ sites, leading to $n=0.125$ and 0.625,
respectively. At any given non-special filling, $\ave{sign}$ deteriorates
steadily as the temperature decreases, rendering the simulations 
inpractical in some cases. Further, Fig.\ \ref{fig1}(b) shows
that for a given temperature, the dip in $\ave{sign}$ gets deeper as the
system size increases. One should note, however, that the position of the
global minimum is roughly independent of the system size, which, for
fillings away from the dip, allows one
to safely monitor size effects on the calculated properties. In this respect, the three-dimensional model is much
less friendly, as shown in Fig.\ \ref{fig1}(c): for $0.3\lesssim n< 1$,
$\ave{sign}$ is never larger than 0.5 at the same filling for both system
sizes.

\begin{figure}
{\centering\resizebox*{3.4in}{!}{\includegraphics*{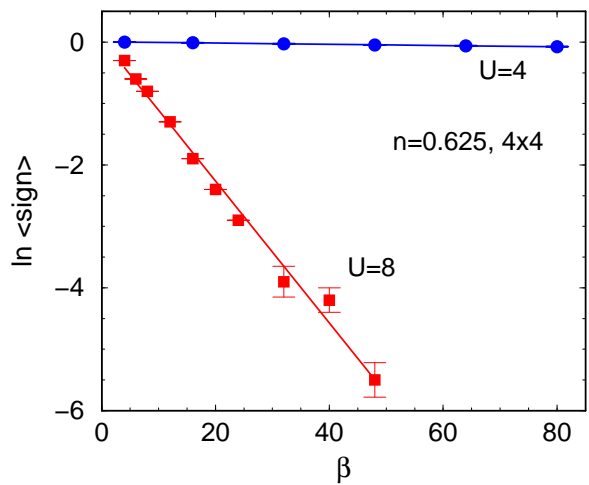}}}
\caption{\label{fig2} 
The logarithm of the average sign of the product of fermionic determinants
as a function of inverse temperature, for the Hubbard model on a $4\times
4$ square lattice with $n=0.625$, and for different values of the Coulomb
repulsion: $U=4$ (circles) and 8 (squares).  Lines are fits through the
data points. Adapted from Ref.\ \cite{Loh90}.}
\end{figure}

It is also instructive to discuss the dependence of $\ave{sign}$ with 
temperature, keeping fixed both the system size and the band filling.  In
Fig.\ \ref{fig2} we show $\ln \ave{sign}$ {\it vs.} $\beta$ for the $4\times4$ lattice at
the closed-shell filling $n=0.625$. (Actually, these data have been
obtained in Ref.\ \cite{Loh90} by means of a ground state algorithm, but
they follow a trend similar to those obtained from the determinantal
algorithm discussed here.)  As $U$ increases, the sign deteriorates even
for this special filling. For other fillings the average sign also
decreases with $U$, and confirms the general expectation \cite{Loh90} that
\begin{equation}
\ave{sign}\sim e^{-\beta N_s \gamma},
\label{sign-exp}
\end{equation}
where $\gamma$ depends on $n$ and
$U$. While for a given $n$, the dependence of $\gamma$ on $U$ is 
monotonic, for a given $U$, $\gamma$ is smaller at the special fillings 
than elsewhere.

The fundamental question then is how to prevent, or at least to minimize,
this minus-sign problem. While one could be tempted to attribute the
presence of negative-weight configurations to the special choice of Hubbard-Stratonovich 
transformations (HST's) used, it has been argued \cite{Batrouni93} that even the
most general forms of HST's are unable to remove the sign
problem. It therefore seems that the problem is of a more fundamental
nature.

\begin{figure}
{\centering\resizebox*{3.4in}{!}{\includegraphics*{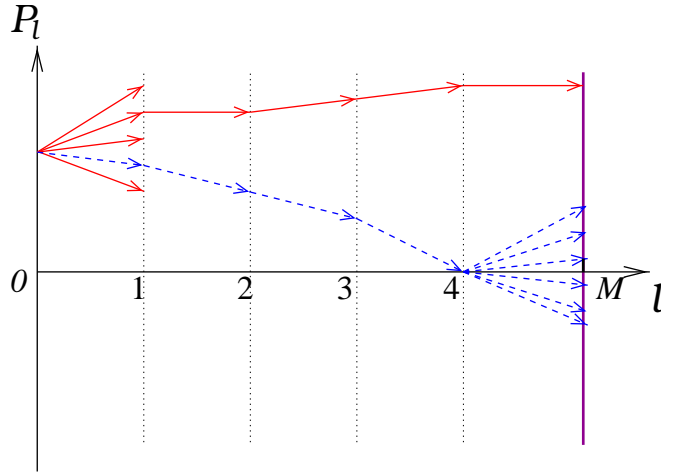}}}
\caption{\label{fig3} 
Schematic behaviour of partial paths (see text) as a function of their
length. Only two representative paths emerging from the ``shower'' at
$\ell=0$ are followed: one (full curve) leads to a positive contribution
when it reaches $\ell=M$, whereas the other (dashed curve) reaches
$\mathcal{P}=0$ at some $\ell_0$.}
\end{figure}

In order to pinpoint the origin of the problem, let us
change the notation slightly and write the partition function as
\begin{equation}
{\mathcal Z}=
%\sum_{\{s(1)\},\{s(2)\},\cdots,\{s(M)\}}
\sum_{S}
\sctr \prod_\s \bb^\s(S_M) \bb^\s(S_{M-1})\cdots \bb^\s(S_1), 
\label{path1}
\end{equation}
where a generic HS configuration of the $\ell$-th time slice is now 
denoted by
$S_\ell\equiv\{s_{\mathbf{1}}(\ell),s_{\mathbf{2}}(\ell),\cdots
s_{\mathbf{N_s}}(\ell)\}$; 
the set $S\equiv \{ S_1,S_2,\cdots S_{M}\}$
defines a \emph{complete path} in the space of HS fields.
Instead of applying the HST to all time slices at once,
one can apply it to each time slice in succession, and collect 
the contribution from a given HS configuration of the first 
$\ell$ time slices into the  
\emph{partial path} \cite{Zhang}, 
%\begin{equation}
\begin{eqnarray}
\mathcal{P}_\ell\left(\{S_1,S_2,\cdots,S_\ell 
\}\right)\qquad\qquad\qquad\qquad\qquad\qquad\qquad&\nonumber\\
\equiv 
\sctr 
   \left[\mathcal{B}\mathcal{B}\cdots\mathcal{B}
   \prod_\s \bb^\s(S_\ell) \bb^\s(S_{\ell-1})\cdots 
\bb^\s(S_1)\right],\ 
\label{partial_path}
%\end{equation}
\end{eqnarray}
where there are $M-\ell$ factors of $\mathcal{B}\equiv e^{-\dt
\mathcal{H}}$ which have not undergone a HST.  Clearly,
$\mathcal{P}_0\equiv \mathcal{Z}$ since it corresponds to the situation in
which no HS fields have been introduced yet; it is therefore positive.  
When HS fields are introduced in the first time slice, a ``shower'' of
$2^{N_s}$ different values of $\mathcal{P}_1$ emerges from
$\mathcal{P}_0$; see Fig.\ \ref{fig3}.  Each subsequent introduction of HS
fields leads to showers of $2^{N_s}$ values of $\mathcal{P}_\ell$ emerging
from each $\mathcal{P}_{\ell-1}$.  In Fig.\ \ref{fig3}, we only follow two
\emph{representative} partial paths: the top one is always positive, while
the bottom one hits the $\ell$-axis at some $\ell_0$. In the latter case,
subsequent showers lead to both negative and positive values of the
partial paths.  According to the framework discussed in Sec.\ \ref{qmc-s},
the simulations are carried out after performing the HST on
all sites of all time slices. In the present context, this amounts to
sampling solely the intersection of all possible paths with the vertical
line at $\ell=M$; see Fig.\ \ref{fig3}.  If one were able to sum over
\emph{all} HS configurations, we would find that the number of positive
$\mathcal{P}_M$'s would exceed that of negative $\mathcal{P}_M$'s by an
amount which, at low temperatures, would be exponentially small. Since in
practice only a limited set of HS configurations is sampled, it is hardly
surprising that we find instances in which configurations leading to
negative weights outnumber those leading to positive weights. This
perspective also helps us to understand why simply discarding those
negative-weighted configurations is not correct: the overall contribution
of the positive-weighted configurations would be overestimated in the
ensemble averages.

The analysis of partial paths is at the heart of a recent proposal
\cite{Zhang} to solve the minus-sign problem. It is based on the fact that
when a partial path touches the $\mathcal{P}=0$ axis, it leads to showers
which, when summed over all subsequent HS fields, give a vanishing
contribution; see Fig.\ \ref{fig3}. In other words, a replacement of all
$\mathcal{B}$'s in Eq.\ (\ref{partial_path}) by
$\sum_{S_{\ell_0+1}\cdots}\prod_\s\bb^\s(S_{\ell_0+1})\cdots$ does not
change the fact that $\mathcal{P}_{\ell_0}=0$. Therefore, if one is able
to follow the `time' evolution of the partial paths, and discard those
that vanish at $\ell<M$, only positive-weight configurations end up
contributing at $\ell=M$. However, though very simple in principle, this
programme is actually quite hard to implement due to the need to handle
the $\mathcal{B}$ factors without using a HS transformation. Zhang
\cite{Zhang} suggested the use of \emph{trial} $\mathcal{B}$'s, and
preliminary results seem encouraging.
Clearly, much more work is needed
along these lines in order to fully assess the efficiency and robustness
of the method.

Likewise, other recent and interesting proposals to deal with the minus
sign problem need to be thoroughly scrutinized. In the meron-cluster
approach \cite{SC99}, HS fields are introduced in all sites as usual, but
during the sampling process (1) the configurations are decomposed into
clusters which can be flipped independently, and (2) a matching between
positive- and negative-weighted configurations is sought; see Ref.\
\cite{SC99} for details, and Ref.\ \cite{Mak98} for another grouping
strategy. Another approach \cite{Asai00}, so far devised for the
ground-state projector algorithm, consists of introducing a
decision-making step to guide walkers to avoid configurations leading to
zero weight; it would be worth examining whether the ideas behind this
adaptive sampling approach could also be applied to the finite temperature
algorithm. In this respect it should be noted that Zhang's approach is
closely related to the Constrained Path Quantum Monte Carlo\cite{CPQMC},
in which the ground-state energy and correlation functions are obtained by
eliminating configurations giving rise to negative overlaps with a trial
state. 

In summary, at the time of this writing, QMC simulations are still plagued
by the negative-sign problem. Many ideas to solve this problem have been
tested over the years, and they require either some bias (through
resorting to trial states) or quite intricate algorithms (which render
simulations of moderately large systems at low temperatures prohibitive in
terms of CPU time), or both. We hope these recent proposals can be
implemented in an optimized way.

\section{Instabilities at low temperatures}
\label{stable}

When the framework discussed so far is implemented into actual
simulations, one faces yet another problem, namely, the fact that the
calculation of Green's functions becomes unstable at low temperatures. As
mentioned in the paragraph below Eq.\ (\ref{gell1}), the Green's functions
can be iterated for about $\tilde\ell\sim 10$ time slices, after which
they have to be calculated from scratch. However, as the temperature is
lowered, {\it i.e.,} for $\beta\gtrsim 4$, $\tilde\ell$ must be decreased
due to large errors in the iterated Green's function (as compared with the
one calculated from scratch). One soon reaches the situation in which the
Green's function has to be calculated from scratch at every time slice
({\it i.e.,} $\tilde\ell=1$).  It should be noted that this corruption
only occurs as one iterates from one time slice to another, and not in the
updating stage within a given time slice \cite{White89}. Worse still is
the fact that as the temperature is lowered further, the Green's function
cannot even be calculated from scratch, since
$\mathsf{O}^\s=\mathsf{1}+\mathsf{A}^\s(\ell)$ 
%[{\it c.f.} Eqs.\ (\ref{gell}) and (\ref{A})] 
becomes so ill-conditioned that it cannot be inverted by simple methods. 
For instance, in two dimensions and for $U=0$, the eigenvalues of $\mathsf{O}^\s$ range between $\sim 1$ and $\sim e^{4\beta}$; 
for $U\neq 0$ we therefore expect the ratio between the largest and 
smaller eigenvalues of $\mathsf{O}^\s$ to grow exponentially with $M$, 
thus becoming singular at low temperatures.

Having located the problem, two solutions have been proposed, which we 
discuss in turn.

\subsection{The Space-time Formulation}
\label{space-time}

The approach used so far can be termed as a \emph{space} formulation,
since the basic ingredient, the Green's function [or, equivalently, the
matrices $\mathsf{A}^\s$ and $\mathsf{O}^\s$, {\it c.f.}\ Eqs.\ 
(\ref{detO}), (\ref{gell}), and (\ref{A})] is 
an $N_s\times N_s$ matrix. 
However, within a field-theoretic framework, if space is discretized 
before integrating out the fermionic degrees of freedom \cite{BSS}, 
the matrices $\mathsf{O}^\s$ are blown up to
\begin{equation}
\mathsf{\hat O}^\s=
\begin{pmatrix}
\mathsf{1} &          0 &         0 & \cdots & 0      &  \bb_M^\s  \\
-\bb_1^\s  & \mathsf{1} &         0 & \cdots & 0      &         0  \\
        0  & -\bb_2^\s  & \mathsf{1}& \cdots & 0      &         0  \\
 \vdots    &  \vdots    &  \vdots   & \cdots & \vdots &  \vdots    \\
        0  &          0 &         0 & \cdots & -\bb_{M-1}^\s& \mathsf{1}\\
\end{pmatrix},
\label{Ohat}
\end{equation}
which is an $(N_sM)\times (N_sM)$ matrix (since each of the $M\times M$
entries is itself an $N_s\times N_s$ matrix); one still has \cite{BSS}
\begin{equation}
\det \mathsf{\hat O}^\s = \det \left[ \mathsf{1} + \bb_M^\s \cdots 
\bb_1^\s\right],
\label{detOhat}
\end{equation}
as in Eq.\ (\ref{detB}), and
\begin{equation}
\mathcal{Z}=\left(\frac{1}{2}\right)^{L^dM}
\underset{\{s\}}{\text{\Large\rm Tr}}\,  
\det {\mathsf{\hat O}}^\up(\{s\})\cdot \det 
{\mathsf{\hat O}}^\dn(\{s\}).
\label{detOhat2}
\end{equation}
Taking the inverse of $\mathsf{\hat O}^\s$ 
yields immediately the space-time Green's function matrix,
\begin{equation}
\mathsf{\hat g}^\s=\left[\mathsf{\hat O}^\s\right]^{-1}.
\label{gOhat}
\end{equation} 

This space-time formulation has the advantage of shrinking 
the range of eigenvalues of $\mathsf{\hat O}^\s$: the ratio between its 
largest and smallest eigenvalues now grows only linearly with $M$, 
thus becoming numerically stable \cite{Hirsch88}. 
Though this approach has been extremely useful in the study of magnetic 
impurities \cite{Fye}, it does slow down the simulations when applied 
to the usual Hubbard model \cite{HirschLin88}. 
Indeed, dealing with $(N_sM)\times (N_sM)$ Green's functions would 
require $(N_sM)^2$ operations per update, $N_s^3M^2$ operations per time 
slice, and, finally, $(N_sM)^3$ operations per sweep of the space-time 
lattice. Sweeping through the space-time lattice with $\mathsf{\hat g}^\s$ 
is then a factor of $M^2$ slower than sweeping with $\mathsf{g}^\s$. 

A solution of compromise between these two formulations was proposed by
Hirsch \cite{Hirsch88}. Instead of using one time slice as a new entry 
in $\mathsf{\hat O}^\s$ [Eq.\ (\ref{Ohat})], we collapse $M_0< M$ 
time slices into a new entry. That is, by taking $M_0\equiv M/p$, with 
$p$ an integer, one now deals with $(N_sp)\times (N_sp)$ matrices of
the form
\begin{widetext}
\begin{equation}
\mathsf{\hat O}_{M_0}^\s(1)=
\begin{pmatrix}
\mathsf{1} & 0 & 0 & \cdots & 0 &  \bb_{pM_0}^\s\cdots\bb_{(p-1)M_0+1}^\s\\
-\bb_{M_0}^\s\bb_{M_0-1}^\s\cdots\bb_1^\s&\mathsf{1}&0&\cdots&0&0  \\
0&-\bb_{2M_0}^\s\cdots\bb_{M_0+1}^\s&\mathsf{1}&\cdots & 0 &    0  \\
 \vdots    &  \vdots    &  \vdots   & \cdots & \vdots &  \vdots    \\
0 & 0 & 0 & \cdots & -\bb_{(p-1)M_0}^\s\cdots\bb_{(p-2)M_0+1}& \mathsf{1}\\
\end{pmatrix},
\label{Ohatp}
\end{equation}
in terms of which the partition function is calculated as in Eq.\
(\ref{detOhat2}). The label 1 of $\mathsf{\hat O}_{M_0}^\s$ indicates that
the product of $\bb$'s start at the first time slice of each of the $p$
groupings. As a consequence, the time-dependent Green's function 
sub-matrices,
$\mathsf{G}^\s(\ell_1;\ell_2)$, connecting the first time slice of each
grouping with either itself or with the first time slice of subsequent
groupings, are readily given by \cite{Hirsch88}
\begin{equation}
\left[\mathsf{\hat O}_{M_0}^\s(1)\right]^{-1}\equiv \mathsf{\hat 
g}_{M_0}^\s(1)=
\begin{pmatrix}
\mathsf{G}(1,1) &\mathsf{G}(1,M_0+1)& \cdots & \mathsf{G}(1,(p-1)M_0+1) \\
\mathsf{G}(M_0+1,1) & \mathsf{G}(M_0+1,M_0+1) & \cdots & 
\mathsf{G}(M_0+1,(p-1)M_0+1)\\
\vdots & \vdots & \cdots & \vdots \\
\mathsf{G}((p-1)M_0+1,1) & \mathsf{G}((p-1)M_0+1,M_0+1) & \cdots & 
\mathsf{G}((p-1)M_0+1,(p-1)M_0+1)\\
\end{pmatrix},
\label{gtmatrix}
\end{equation}
\end{widetext}
where the spin indices in $\mathsf{G}$ have been omitted for 
simplicity.
The Green's functions connecting the $\ell$-th time slices of the $p$
groupings are similarly obtained from the inversion of $\mathsf{\hat
O}_{M_0}(\ell)$, which, in turn, is obtained by increasing all time
indices of the $\bb$'s in Eq.\ (\ref{Ohatp}) by $\ell-1$.

In the course of simulations, one starts with the time-expanded
Green's function calculated from scratch through Eq.\ (\ref{gtmatrix}),
and sweeps all sites in the first time slice of each of the $p$ groupings.
Each time the move is accepted, the Green's function at that time slice is
updated according to Eq.\ (\ref{newg}). After sweeping over these sets of
lattice sites, one iterates the
Green's functions to obtain the elements of $\mathsf{\hat g}_{M_0}(2)$
through \cite{Hirsch88}
\begin{equation}
\mathsf{G}(\ell_1+1,\ell_2+1)= 
\bb_{\ell_1}\mathsf{G}(\ell_1,\ell_2)\bb_{\ell_2}^{-1}.
\label{gtupdate}
\end{equation}
We then sweep through all sites of the second time slice of each of the
$p$ groupings, and so forth, until all time slices are swept. Note that
since Eq.\ (\ref{gtupdate}) is only used $M_0$ times (as opposed to $M$
times) it does not lead to instabilities for $M_0$ small enough.

Each Green's function update with Eq.\ (\ref{newg}) requires
$\sim(N_sp)^2$ operations;  sweeping through all spatial sites of one time
slice in each of every $p$ groupings therefore requires $\sim(N_sp)^3$
operations. Since there are $M_0$ slices on each grouping, one has,
finally, a total of $\sim N_s^3Mp^2$ operations per sweep, which sets the
scale of computer time; this should be compared with $N_s^3M$ for the
original implementation, and $(N_sM)^3$ for the impurity algorithm. The
strategy is then to keep $M_0\sim 20$ and let $p$ increase as the
temperature is lowered. With this algorithm, values of $\beta \sim 20-30$ 
have been achieved in many studies of the Hubbard model; see, e.g., Refs.\
\cite{HirschTang89,dS89,SWZ,dS93,dS94,dS92,dS95,Ghosh,tclp00}. It should also
be mentioned that since the unequal-time Green's function is calculated at
each step, this space-time formulation is especially convenient when one
needs frequency-dependent quantities \cite{SWZ}.

\subsection{Matrix-decomposition stabilization}
\label{matrix-decomp}

Let us assume that $M_0$ of the $\bb$ matrices can be multiplied without 
deteriorating the accuracy. One then defines \cite{White89,Loh92}
\begin{equation}
\mathsf{\tilde A}_1^\s(\ell)\equiv \bb_{\ell+M_0}^\s \bb_{\ell+M_0-1}^\s 
\cdots 
\bb_{\ell}^\s,
\label{P1}
\end{equation}
which, by Gram-Schmidt orthogonalization, can be decomposed into
\begin{equation}
\mathsf{\tilde 
A}_1^\s(\ell)=\mathsf{U}_1^\s\mathsf{D}_1^\s\mathsf{R}_1^\s,
\label{P-decomp}
\end{equation}
where $\mathsf{U}_1^\s$ is a well-conditioned orthogonal matrix,
$\mathsf{D}_1^\s$ is a diagonal matrix with a wide spectrum (in the sense
discussed at the beginning of the Section), and $\mathsf{R}_1^\s$ is a
right triangular matrix which turns out to be well-conditioned
\cite{White89}.

Using the fact that $\mathsf{U}_1^\s$ is well-conditioned, we can multiply 
it by another group of $M_0$ matrices,
\begin{equation}
\mathsf{Q}=\bb_{\ell+2M_0}^\s \bb_{\ell+2M_0-1}^\s \cdots 
\bb_{\ell+M_0+1}^\s\mathsf{U}_1^\s,
\label{Q}
\end{equation}
without compromising the accuracy. We then form the product 
\begin{equation}
\mathsf{Q}^\prime = \mathsf{Q}  \mathsf{D}_1^\s,
\label{Qprime}
\end{equation} 
which amounts to a simple rescaling of the columns, without jeopardizing 
the stability for the subsequent decomposition as
\begin{equation}
\mathsf{Q}^\prime = \mathsf{U}_2^\s\mathsf{D}_2^\s\mathsf{\tilde R}_2^\s,
\label{Q-decomp}
\end{equation}
where the matrices $\mathsf{U}_2^\s$, $\mathsf{D}_2^\s$, and
$\mathsf{\tilde R}_2^\s$ satisfy the same conditions as in the first step,
Eq.\ (\ref{P-decomp}). With $\mathsf{R}_2^\s\equiv\mathsf{\tilde
R}_2^\s\mathsf{R}_2^\s$, we conclude the second step with
\begin{equation}
\mathsf{\tilde A}_2^\s(\ell) = 
\mathsf{U}_2^\s\mathsf{D}_2^\s\mathsf{R}_2^\s.
\label{P2}
\end{equation}
This process is then repeated $p=M/M_0$ times, to finally recover Eq.\ 
(\ref{A}), in the form \cite{White89}
\begin{equation}
\mathsf{A}^\s(\ell)=\mathsf{\tilde A}_{p}^\s=
\mathsf{U}_{p}^\s\mathsf{D}_{p}^\s\mathsf{R}_{p}^\s.
\label{A-decomp}
\end{equation}

The equal-time Green's function is therefore calculated through Eq.\ 
(\ref{gell}), as before, but care must be taken when adding the identity 
matrix to $\mathsf{A}$, since the latter involves 
the wide-spectrum matrix $\mathsf{D}_{p}$. 
One therefore writes
\begin{equation}
\mathsf{1}=\mathsf{U}_{p}^\s 
\left[\mathsf{U}_{p}^\s\right]^{-1}
\mathsf{R}_{p}^\s
\left[\mathsf{R}_{p}^\s\right]^{-1},
\end{equation}
so that the inverse of Eq.\ (\ref{gell}) becomes
\begin{equation}
\left[\mathsf{g}^\s(\ell)\right]^{-1}=
 \mathsf{U}_{p}^\s \mathsf{P}^\s
\mathsf{R}_{p}^\s,
\end{equation}
with 
\begin{equation}
\mathsf{P}^\s = \left[\mathsf{U}_{p}^\s\right]^{-1} 
\left[\mathsf{R}_{p}^\s\right]^{-1} + \mathsf{D}_{p}^\s.
\label{Paux}
\end{equation}
We now apply another decomposition to $\mathsf{P}$, the result of which is 
multiplied to the left by $\mathsf{U}_{p}^\s$, and to the right by  
$\mathsf{R}_{p}^\s$, in order to express the Green's function in the form 
\cite{White89}
\begin{equation}
\left[\mathsf{g}^\s(\ell)\right]^{-1}= \mathsf{U}^\s \mathsf{D}^\s   
\mathsf{R}^\s.
\label{g-decomp}
\end{equation}

In the course of simulations, one proceeds exactly as in Sec.\
\ref{qmc-s}, by updating the Green's function through iterations, which takes
up $\sim N_s^3M$ operations. Again, the iteration from one time slice to
another is limited to about $\tilde\ell$ time slices, and it turns out
that a significant fraction of the computer time is spent in calculating
the Green's function from scratch. Indeed, a fresh $\mathsf{g}$ is 
calculated $M/{\tilde\ell}$ times, each of which involving $p$
decompositions, each of which taking about $N_s^3$ operations. 
Therefore, taking ${\tilde\ell}\sim p$,
the dominant scale of computer time is $\sim N_s^3 p^2$ instead, which
is about a factor of $M$ faster than the space-time algorithm, though
without the bonus of the unequal-time Green's functions. This
matrix decomposition algorithm has also been efficiently applied 
to several studies of the Hubbard model, and
values of $\beta \sim 20 - 30$ have been easily achieved; see, {\it e.g.,}
Refs.\
\cite{White89,Moreo90,Loh90,Moreo91a,Moreo91,SWZ,Trivedi96,Scalettar99,
Paiva01}.

\section{Conclusions}
\label{conc}

We have reviewed the Determinantal Quantum Monte Carlo technique for
fermionic problems. Since the seminal proposal by Blankenbecler,
Scalapino, and Sugar \cite{BSS}, over twenty years ago, this method has
evolved tremendously. Stabilization techniques allowed the calculation of
a variety of quantities at very low temperatures, but the minus-sign
problem still plagues the simulations, restricting a complete analysis
over a wide range of band fillings and coupling constants. In this
respect, it should be mentioned that other implementations of QMC, such as
ground-state algorithms (see Ref.\ \cite{vdl92}), also suffer from this
minus-sign problem. An efficient solution to this problem would be a major
contribution to the field. 

Most of the discussion was centred in the simple Hubbard model, but 
advances have been made in the QMC study of other models, such as the
extended Hubbard model \cite{Callaway}, the Anderson impurity model 
\cite{Fye}, and the Kondo lattice model \cite{Assaad}. The first 
applications of QMC simulations to disordered systems have appeared
only recently; see, {\it e.g.,} Refs.\ \cite{Trivedi96,Scalettar99}.
With the ever increasing power of personal computers and workstations,
one can foresee that many properties of these and of more ellaborate 
and realistic models will soon be elucidated.

\begin{acknowledgments}

The author is grateful to F.\ C.\ Alcaraz and S.\ R.\ A.\ Salinas for the
invitation to lecture at the 2002 Brazilian School on Statistical
Mechanics, and for the constant encouragement to write these notes.  
Financial support from the Brazilian agencies CNPq, FAPERJ, and
Mi\-nis\-t\'erio de Ci\^encia e Tecnologia, through the Millenium
Institute for Nanosciences, is also gratefully acknowledged.

\end{acknowledgments}

\appendix

\section{Grouping products of exponentials}
\label{group}

One frequently needs to group products of exponentials into a single
exponential. Here we establish a crucial result for our
current purposes: the product of bilinear forms can be grouped into a
single exponential of a bilinear form.

To show this result \cite{vdl92}, let the generic Hamiltonian be expressed
as (we omit spin indices)
\begin{equation}
\mathcal{H}=\sum_{i,j} c_i^{\dagger} h_{ij} 
c_j^{\phantom{\dagger}},
\label{H-ell}
\end{equation}
where $\mathsf{h}$ is a matrix representation of the 
operator $\mathcal{H}$. The 
`time'
evolution of Heisenberg operators [{\it c.f.,} Eq.\ (\ref{heispic})],
satisfies a differential equation, whose solution can be found to be
\begin{equation}
c_i^{\dagger}(\tau) = 
\sum_i \left[ e^{-\tau \mathsf{h}}\right]_{ij} c_i^{\dagger}.
\label{c_dagg-tau}
\end{equation}
If we now take the linear combination
\begin{equation}
c_\alpha^\dagger=\sum_i B_{\alpha i}\ c_i^\dagger,
\label{c_a-d}
\end{equation}
we see that 
\begin{eqnarray}
c_\alpha^\dagger(\dt) &=&
 \sum_i \sum_j \left[ e^{-\dt \mathsf{h}}\right]_{ij} B_{\alpha j}\ 
c_i^\dagger=\nonumber\\
&\equiv & \sum_i {\tilde B}_{\alpha i}\ c_i^\dagger,
\end{eqnarray}
where the last passage has been used as the definition of ${\tilde B}$.

Now we introduce the product over time slices, through
%If we further define 
\begin{equation}
\mathcal{U}\equiv \prod_\ell e^{-\dt \mathcal{H}(\ell)}, 
\end{equation}
so that the $\mathsf{h}$'s now acquire $\ell$-labels, and we have
\begin{eqnarray}
\mathcal{U} c_\alpha^\dagger &=& 
   \sum_i \sum_j \left[ \prod_\ell e^{-\dt \mathsf{h}(\ell)}\right]_{ji} 
B_{\alpha j}\ c_i^\dagger \mathcal{U}=\nonumber\\
&\equiv& \sum_i \sum_j \left[e^{-\dt {\mathsf{\tilde h}}}\right]_{ji}
B_{\alpha j}\ c_i^\dagger \mathcal{U},
\end{eqnarray}
where, again, the last passage has been used to define $\mathsf{\tilde 
h}$. We can then write 
\begin{equation}
\mathcal{U} c_\alpha^\dagger \mathcal{U}^{-1}=\sum_i \sum_j \left[e^{-\dt 
{\mathsf{\tilde h}}}\right]_{ji}
B_{\alpha j}\ c_i^\dagger,
\end{equation}
so that $\mathcal{U}$ indeed behaves as a single exponential of the 
effective bilinear Hamiltonian $\mathcal{\tilde H}$, whose elements 
are defined by
\begin{equation}
\left[ \prod_\ell e^{-\dt \mathsf{h}(\ell)}\right]_{ij}=
\left[ e^{-\dt \mathsf{\tilde h}}\right]_{ij}.
\end{equation}

\section{Tracing out exponentials of fermion operators}
\label{trace}

We now use the results of App.\ \ref{group} to show that the trace over
the fermionic degrees of freedom can be expressed in the form of a 
determinant. Given that $\mathcal{\tilde H}$ is bilinear in fermion 
operators, it can be straightforwardly 
diagonalized\cite{BSS,Hirsch85,vdl92,Loh92}:
\begin{equation}
\mathcal{\tilde H}=\sum_\alpha \varepsilon_\alpha c_\alpha^\dagger 
c_\alpha^{\phantom{\dagger}},
\end{equation}
where, for completeness, the relations between `old' and `new' fermion 
operators are
\begin{equation}
c_\alpha=\sum_i \braket{\alpha}{i}c_i,
\label{c-a}
\end{equation}
and
\begin{equation}
c_\alpha^\dagger=\sum_i \braket{i}{\alpha}c_i^\dagger;
\label{c-a-dag}
\end{equation}
the inverse relations are
\begin{equation}
c_i=\sum_\alpha \braket{i}{\alpha}c_\alpha,
\label{c-i}
\end{equation}
and 
\begin{equation}
c_i^\dagger=\sum_\alpha \braket{\alpha}{i}c_\alpha^\dagger.
\label{c-i-dag}
\end{equation}

Given the diagonal form, the fermion trace is then immediately given by
\begin{eqnarray}
\sctr \prod_\ell e^{-\dt \mathcal{H}(\ell)}& = & \sctr e^{-\dt 
\mathcal{\tilde H}}=\nonumber\\
&=&\sctr e^{-\dt \sum_\alpha \varepsilon_\alpha 
c_\alpha^\dagger c_\alpha^{\phantom{\dagger}}}=\nonumber\\
&=&\sctr \prod_\alpha e^{-\dt \varepsilon_\alpha
c_\alpha^\dagger c_\alpha^{\phantom{\dagger}}}=\nonumber\\ 
&=& \prod_\alpha \left(1+e^{-\dt \varepsilon_\alpha}\right)=\nonumber\\
&=& \det \left[ \mathsf{1}+ e^{-\dt \mathcal{\tilde H}}\right]=\nonumber\\
&=& \det \left[ \mathsf{1}+ \prod_\ell e^{-\dt \mathsf{h}(\ell)}\right].
\label{trace-det}
\end{eqnarray}

\section{Equal-time Green's functions}
\label{etgf}

The equal-time Green's function, defined as the fermion 
average for a given configuration of the HS fields, is 
(we omit the fermion spin label) 
\begin{equation}
\have{c_i^{\phantom{\dagger}}c_j^{\dagger}}=
\frac{1}{\mathcal{Z}}\sctr\ e^{-\beta\mathcal{H}} 
c_i^{\phantom{\dagger}}c_j^\dagger,
\label{gdef}
\end{equation}
where one assumes, for the time being, that $c_i^{\phantom{\dagger}}$ 
and $c_j^{\dagger}$ carry no time label.
  
As in the calculation of the partition function, Eq.\ (\ref{Znew}), 
we write $\exp(-\beta\mathcal{H})$ as a product of $M$ factors
$\exp(-\dt\mathcal{H})$. We then introduce the matrix representation,
\begin{equation}
D_\ell = 
e^{-\dt\sum_{i,j}c_{i}^{\dagger} K_{ij} 
c_{j}^{\phantom{\dagger}}}
e^{-\dt\sum_{i}c_{i}^{\dagger} V_{i}(\ell) 
c_{i}^{\phantom{\dagger}}},
\label{D-deff}
\end{equation}
to write
\begin{equation}
\have{c_i^{\phantom{\dagger}}c_j^{\dagger}}=\frac{\sctr D_M\cdots D_1 
c_i^{\phantom{\dagger}}c_j^{\dagger}}
{\sctr D_M\cdots D_1}.
\label{gcalc1}
\end{equation}
In order to evaluate $\sctr$, we first change to the diagonal 
representation through Eqs.\ (\ref{c-i}) and (\ref{c-i-dag}), 
to write
\begin{equation}
\have{c_i^{\phantom{\dagger}}c_j^{\dagger}}=\sum_{\alpha',\alpha''}\braket{i}{\alpha'}\braket{\alpha''}{j}\ 
\frac{\sctr c_{\alpha'}^{\phantom{\dagger}}c_{\alpha''}^{\dagger} D_M\cdots D_1}
{\sctr D_M\cdots D_1}.
\label{gcalc2}
\end{equation}
When we take the $\sctr$ on the diagonal basis, $c_{\alpha'}^{\phantom{\dagger}}c_{\alpha''}^{\dagger}$ act on the bra to the
left, so a nonzero contribution only occurs for $\alpha'=\alpha''$.
Then, all states but $|\alpha'\rangle$ contribute to the product 
with $\left(1+\exp(-\dt\varepsilon_\alpha)\right)$ [see Eq.\ (\ref{trace-det})]
which cancels with the same term in the denominator. We are then
left with
\begin{eqnarray}
\have{c_i^{\phantom{\dagger}}c_j^{\dagger}}&=&\sum_{\alpha'} |\alpha'\rangle \langle i | 
\frac{1}{1+e^{-\dt\varepsilon_\alpha'}}
|j\rangle \langle \alpha'| =\nonumber\\
&=&\left[ \frac{1}{\mathsf{1}+\bb_M\cdots\bb_1}\right]_{ij}=[\mathsf{g}]_{ij}.
\label{gcalc3}
\end{eqnarray}

We can generalize the above result and write the Green's function at the 
$\ell$-th time slice as
\begin{equation}
\mathsf{g}^\s(\ell)\equiv 
\left[ \mathsf{1} + 
\bb_{\ell-1}\bb_{\ell-2}\ldots
\bb_{1}\bb_{M}\ldots\bb_{\ell}
\right]^{-1}.
\label{gell2}
\end{equation}
The reader should convince himself/herself that the average of 
$\mathsf{g}(\ell)$ over \emph{all} HS configurations should be
independent of $\ell$.

\section{Unequal-time Green's functions}
\label{utgf}

The unequal-time Green's functions are defined as (we omit the spin 
indices, for simplicity)
\begin{widetext}
\begin{equation}
G_{ij}(\ell_1;\ell_2)\equiv\have{c_i^{\phantom{\dagger}}(\ell_1)c_j^\dagger(\ell_2)}=
\frac{1}{\mathcal{Z}}\mathcal{T}\!r \, 
e^{-\beta\mathcal{H}}e^{\ell_1\dt\mathcal{H}} 
c_i^{\phantom{\dagger}}
e^{-\ell_1\dt\mathcal{H}}e^{\ell_2\dt\mathcal{H}}
c_j^\dagger e^{-\ell_2\dt\mathcal{H}}.
\label{utgfdef}
\end{equation} 

As in the calculation of the partition function, we write the first
exponential to the right of the $\sctr$ as a product of $M$ factors
$e^{-\dt\mathcal{H}}$, so that there are, in effect, $M-\ell_1$ of these
factors before $c_i$; by the same token, there are $\ell_1-\ell_2$ such
factors between $c_i^{\phantom{\dagger}}$ and $c_j^\dagger$, and $\ell_2$
factors to the right of $c_j^\dagger$. We then use the cyclic property of
the $\sctr$ to bring these latter factors to the front, and introduce
$\ell_1-\ell_2$ factors of $e^{-\dt\mathcal{H}}$ followed by the same
number of $e^{\dt\mathcal{H}}$ factors before $c_i$. The Suzuki-Trotter
decomposition is then used to break the kinetic energy and interaction
terms, and the discrete HS transformation is applied to the exponentials
involving the interaction; this adds a time-slice label to the resulting
group of exponentials, and we can write
\begin{equation}
%\have{c_i^{\phantom{\dagger}}(\ell_1)c_j^\dagger(\ell_2)}=
G_{ij}(\ell_1;\ell_2)=
\frac{1}{\mathcal{Z}}\mathcal{T}\!r \, 
D_{\ell_2}\cdots D_1D_M\cdots D_{\ell+1}D_{\ell}\cdots 
D_{\ell_2+1} 
\left[D_{\ell_1}\cdots D_{\ell_2+1}\right]^{-1} 
c_i^{\phantom{\dagger}}
\left[D_{\ell_1}\cdots D_{\ell_2+1}\right]
c_j^\dagger,
\label{utD}
\end{equation}
%\end{widetext} 
where $D_\ell$ is given by Eq.\ (\ref{D-deff}).
The product of $D$'s can be expressed in its diagonal 
representation, $\{|\alpha\rangle\}$, 
\begin{equation}
D_{\ell_1}\cdots D_{\ell_2+1}
=e^{\sum_\alpha |\alpha\rangle \ep_\alpha \langle\alpha|},
\label{D-alpha}
\end{equation}
which can also be used to express $c_i$, as in Eq.\ (\ref{c-i}). 
Since
%\begin{widetext} 
\begin{equation}
c_\gamma\, e^{-\ep_\al c_\al^\dagger c_\al^{\phantom{\dagger}}}=
\begin{cases}
e^{-\ep_\gamma}c_\gamma 
       &\text{if $\al=\gamma$},\\
e^{-\ep_\al c_\al^\dagger c_\al^{\phantom{\dagger}}}\, c_\gamma
       &\text{if $\al\neq\gamma$,}
\end{cases} 
\label{comm}
\end{equation}
we have
\begin{equation}
\sum_\gamma\, \langle i|\gamma\rangle 
\left[D_{\ell_1}\cdots D_{\ell_2+1}\right]^{-1}
c_\gamma
\left[D_{\ell_1}\cdots D_{\ell_2+1}\right]
=
\sum_\gamma\ \langle i|\gamma\rangle\  
e^{-\ep_\gamma(1-n_\gamma)}c_\gamma.
\label{mid01}
\end{equation}
Thus, since $1-n_\al=0,1$, we can write
\begin{equation}
e^{-\ep_\gamma(1- n_\gamma)}=
\langle\gamma|\bb_{\ell_1}\cdots\bb_{\ell_2+1}|\gamma\rangle,
\label{mid02}
\end{equation}
which leads to
\begin{equation}
\left[D_{\ell_1}\cdots D_{\ell_2+1}\right]^{-1}
c_i
\left[D_{\ell_1}\cdots D_{\ell_2+1}\right]
=\sum_k  
\, \langle i|\bb_{\ell_1}\cdots\bb_{\ell_2+1}|k\rangle
\ c_k,
\label{matrixprod}
\end{equation}
where the sum is in site indices.

With (\ref{matrixprod}), Eq.\ (\ref{utD}) becomes
\begin{equation}
%\have{c_i^{\phantom{\dagger}}(\ell_1)c_j^\dagger(\ell_2)}=
G_{ij}(\ell_1;\ell_2)=
\frac{1}{\mathcal{Z}}
\sum_k
\, \langle i|\bb_{\ell_1}\cdots\bb_{\ell_2+1}|k\rangle
\ \mathcal{T}\!r \, 
D_{\ell_2}\cdots D_1D_M\cdots D_{\ell_2+1}
c_k^{\phantom{\dagger}}c_j^\dagger, 
\label{utD02}
\end{equation}
\end{widetext} 
which can, finally, be expressed in terms of the equal-time Green's 
functions, 
Eq.\ (\ref{gell2}), as
\begin{equation}
%\have{c_i^{\phantom{\dagger}}(\ell_1)c_{j}^{\dagger}(\ell_2)}=
G_{ij}(\ell_1;\ell_2)=
\left[\bb_{\ell_1}\bb_{\ell_1-1}\ldots \bb_{\ell_2+1}\,
\mathsf{g}(\ell_2+1)\right]_{ij},
\label{gtau-app}
\end{equation}

The reader should check that an analogous calculation, still for 
$\ell_1>\ell_2$, leads to
\begin{eqnarray}
&\tilde G_{ij}(\ell_1;\ell_2)
&\equiv
\have{c_i^\dagger(\ell_1)c_j^{\phantom{\dagger}}(\ell_2)}
=\nonumber\\
&&\Big[\big[\mathsf{1}-\mathsf{g}(\ell_2+1)\big]
\left(\bb_{\ell_1}\bb_{\ell_1-1}\ldots \bb_{\ell_2+1}\right)^{-1}
\Big]_{ij}.\nonumber\\
&& 
\label{gtau2-app}
\end{eqnarray}

In the main text, the Green's functions given in Eqs.\ (\ref{gtau-app}) 
and (\ref{gtau2-app}) acquire a spin index.

\end{document}